\documentstyle[12pt,psfig]{article} 
\newcommand{\rf}[1]{(\ref{#1})}
\newcommand{\beq}{\begin{equation}}
\newcommand{\eeq}{\end{equation}}
\newcommand{\bea}{\begin{eqnarray}}
\newcommand{\eea}{\end{eqnarray}}

\renewcommand{\l}{\lambda}

\renewcommand{\a}{\alpha}

\newcommand{\m}{\mu}
\newcommand{\del}{\delta}

\newcommand{\ep}{\varepsilon}

\newcommand{\k}{\kappa}

\newcommand{\oh}{\frac{1}{2}}
\newcommand{\oq}{\frac{1}{4}}

\newcommand{\dg}{\dagger}

\newcommand{\prt}{\partial}

\newcommand{\cH}{{\cal H}}

\newcommand{\nn}{\nonumber \\}
\newcommand{\mbar}[1]{\overline {#1} \hskip 1pt{}}
\newcommand{\bra}[1]{ \langle #1 | }
\newcommand{\ket}[1]{ | #1 \rangle }
\newcommand{\vac}{ \bra{{\rm vac}} }
\newcommand{\cuum}{ \ket{{\rm vac}} }

\newcommand{\Hj}{{\cH^\star}}
\newcommand{\tHj}{\tilde{\cH}^{\star}}
\newcommand{\tcH}{\tilde{\cH}}
\newcommand{\Hdisk}{\cH_{\rm disk}}
\newcommand{\Hone}{{\mbar \cH}}

\newcommand{\cc}{\mu} 
\newcommand{\T}{T} 
\newcommand{\define}{ \stackrel{\rm def}{\equiv} }

\newcommand{\intdzeta}{ { \int\limits_{- i \infty}^{+ i \infty}
                        \frac{d \zeta}{2 \pi i} } }
\newcommand{\intz}{ { \oint\!\frac{dz}{2 \pi i} } }
\newcommand{\prtz}{ { \frac{\prt}{\prt z} } }

\newcommand{\dj}[1]{ \frac{\prt}{\prt j_{#1}} }
\newcommand{\bJ}{{\mbar W}^{(1)}}
\newcommand{\bW}{\mbar W}

\newcommand{\tJ}{\a}
\newcommand{\rW}[1]{\mbar W^{(#1)'}}
\newcommand{\const}{{\rm const.}}

\newcommand{\Norder}{{}^\circ_\circ}
\newcommand{\dbl}[1]{#1}

\begin{document}
\topmargin 0pt
\oddsidemargin 5mm                              
\headheight 0pt
\headsep 0pt
\topskip 9mm

\hfill    NBI-HE-96-17

\hfill    TIT/HEP--327

\hfill April 1996

\begin{center}
  \vspace{24pt}
  {\large \bf Non-critical string field theory for 2d quantum gravity \\
    coupled to (p,q)--conformal fields }

  \vspace{24pt}

  {\sl J. Ambj\o rn }

  \vspace{6pt}

  The Niels Bohr Institute\\
  Blegdamsvej 17, DK-2100 Copenhagen \O , Denmark\\

  \vspace{12pt}

  and

  \vspace{12pt}

  {\sl Y. Watabiki}

  \vspace{6pt}

  Department of Physics\\
  Tokyo Institute of Technology\\
  Oh-okayama, Meguro, Tokyo 152, Japan\\

\end{center}
\vspace{24pt}

\vfill

\begin{center}
  {\bf Abstract}
\end{center}

\vspace{12pt}

\noindent
We propose a non-critical string field theory for $2d$ quantum gravity 
coupled to ($p$,$q$) conformal fields. 
The Hamiltonian is described by the generators of the $W_p$ algebra, 
and the Schwinger-Dyson equation is equivalent to 
a vacuum condition imposed on the generators of $W_p$ algebra. 

\vfill

\vspace{36pt}



\newpage

\section{Introduction}

Non-critical string theories in dimensions $d <1$, or equivalently,
two-dimensional quantum gravity coupled to conformal theories
with $c <1$, have been intensively analyzed in the last eight years,
either by the use of Liouville field theory or matrix models. 
The  gravity aspects of the theories which refer to metric 
properties have been difficult to handle by continuum methods. 
In the framework of dynamical triangulations 
important progress was made by the calculation of the {\it transfer matrix}
\cite{transfer}.
It allowed the calculation of the two-loop amplitude as a function of
the {\it geodesic distance}. 
In fact, the transfer matrix offered a natural Hamiltonian $\Hone$ 
for the one-string propagation and
the geodesic distance between the two string loops played
the role of {\it proper time} in a Hamiltonian framework.
In \cite{ik}, \cite{jr} and \cite{watabiki1} 
the Hamiltonian $\Hone$ for the one-string propagation 
was generalized to a genuine string field theory 
which allowed the calculation of any string amplitude,
in \cite{ik} from a formal continuum point of view, 
in \cite{jr} from a stochastic quantization point of view, and
in \cite{watabiki1} from the point of view of dynamical triangulations 
which provide an explicit regularization of the theory. 
In \cite{watabiki1} it was in addition  shown that there was universality: 
the details of the random graphs used in the dynamical triangulation 
were not important.
The concept of a string field Hamiltonian 
in the case of $(p,q)$ conformal matter coupled to 
two-dimensional quantum gravity was further developed 
in a series of papers \cite{ik2,iikmns,ikehara,ns}. 
One outcome was that 
in the case of $(p,q) = (m,m+1)$ conformal theories, i.e.
the unitary theories, consistency demanded a proper time $T$ of dimension
$\dim T = 1/m$, if we define the world-sheet to have dimension 2.
In particular, 
this formula shows that the dimension of proper time goes to zero as
the central charge $c= 1-6/m(m+1)$ goes to one for $m\to \infty$.
If we still assume that $T$ can be identified with the geodesic
distance this implies that the internal Hausdorff dimension $d_h \to \infty$.

The concept of geodesic distance in quantum gravity was further
clarified in \cite{aw,kawaietal}, 
where it was shown how the dimension of geodesic distance 
can be extracted unambiguously from the two-point function
by the use of dynamical triangulations in the case of
unitary theories coupled to gravity. 
The two-point function could be calculated explicitly 
in the case of pure gravity, 
i.e. for a $(p,q)=(2,3)$ theory, where $c=0$.  
However, subsequent numerical simulations indicated that 
the geodesic distance has dimension $1/2$
for {\it all} conformal field theories with $0 \leq c \leq 1$ 
coupled quantum gravity \cite{syracuse,ajw}. 
This suggests that the relation between
$T$ and geodesic distance in general might be more complicated than 
anticipated from the study of pure two-dimensional quantum gravity.

As a step towards the  resolution of these problems we present here
detailed study of the Hamiltonian used in string field theory (SFT).
A number of new features appear which have an independent interest,
and we will present these in the rest of this article.

\section{Pure gravity}

In this section we investigate the (2,3)-SFT, i.e. pure gravity,
using a new mode expansion. In addition we will clarify the
relation between the Schwinger-Dyson equations
in the string field theory and the vacuum conditions of the $W_2$ algebra.

\subsection{String field theory for disk topology}

\begin{figure}
  \centerline{\hbox{\psfig{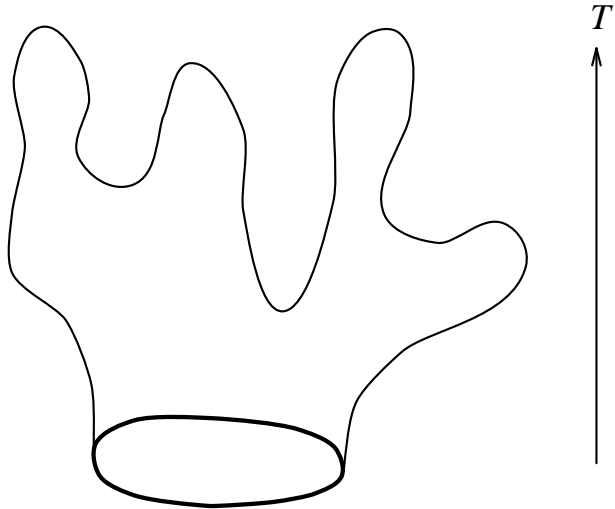}}}
  \caption[fig1]{The propagation in proper time 
    of a closed string which annihilates into 
    the vacuum, allowing only disk topology.}
  \label{fig1}
\end{figure}

Recall the string field theory as formulated in \cite{ik}
in the case where the topology of the two-dimensional
manifold is that of a sphere with boundaries (closed strings).
Let $\Psi^{\dagger}(L)$ and $\Psi(L)$ be operators
which creates and annihilates one closed string
with length $L$, respectively.
The commutation relation of the string operators are 
\beq\label{CommutePsi}
[  \Psi (L)  ,  \Psi^{\dagger} (L')  ]   = \delta ( L - L' ) , 
\eeq
\beq\label{CommutePsi1}
[ \Psi^{\dagger} (L)  ,  \Psi^{\dagger }(L')  ]  = 
[ \Psi (L)  , \Psi (L') ] =  0 .
\eeq
In the Hamiltonian formalism the disk amplitude,
$F_1^{(0)}(L;\cc)$, i.e. the amplitude that one closed
string annihilates into the vacuum, is  obtained by
\beq\label{DiskAmp1}
F_1^{(0)} (L;\cc) \ = \
\lim_{\T \rightarrow \infty}
\vac e^{- \T \Hdisk} \Psi^\dagger (L) \cuum ,
\eeq
where $\T$ is the so-called proper time, $\m$ the cosmological constant and
$\Hdisk$  the Hamiltonian for disk topology \cite{ik}.
In Fig.\ \ref{fig1} we show a typical configuration
which contributes to the disk amplitude defined in \rf{DiskAmp1}.
It should be noted that
\beq\label{VacuumCondition}
\Hdisk \cuum  \ = \  0 ,
\eeq
is a necessary condition in this formalism.
The condition \rf{VacuumCondition} means that
a string is not created from the vacuum, i.e. the
stability of the vacuum against decay into another physical state.
The Laplace transformation of $F_1^{(0)}(L;\cc)$ is given by
\bea
F_1^{(0)} (\zeta;\cc)
&=& \int_0^\infty \! dL \, e^{-\zeta L} F_1^{(0)} (L;\cc)
\nonumber\\
&=& \lambda (\zeta)  \, +  \, F_1^{{\rm univ}(0)} (\zeta;\cc) \, ,
\label{DiskAmp2}
\eea
where $\lambda(\zeta)$ and $F_1^{{\rm univ}(0)} (\zeta;\cc)$
are a divergent non-universal part and a universal part, respectively,
given by,
\beq\label{Lambda}
\lambda (\zeta)
= ({\rm const.}) \times \ep^{-3/2}
\, - \, ({\rm const.}) \times \ep^{-1/2} \zeta \, ,
\eeq
\beq\label{DiskUniv}
F_1^{{\rm univ}(0)} (\zeta;\cc)
= ( \zeta - \frac{\sqrt{\cc}}{2} ) \sqrt{ \zeta + \sqrt{\cc} } \, ,
\eeq
where $\ep$ is a cut off. 
In the context of dynamical
triangulations we can view $\ep$ as the lattice spacing, i.e. the
link length in the triangulations.

As already mentioned the above formalism is somewhat singular, 
and if we try to express $\Hdisk$ in terms of $\Psi(L)$ and $\Psi^\dg(L)$
even the simplest commutators will need a regularization.
For example, $\int_0^L dL' F_1^{(0)}(L';\cc) F_1^{(0)}(L-L';\cc)$,
which is obtained from $[ \Hdisk , \Psi^\dagger (L) ]$,
suffers from such a divergence.   
This problem was analyzed carefully in \cite{watabiki1}, where it was
shown that a subtraction of a singular part of the Laplace transform
of string wave function $\Psi^\dg(L)$ would
render the expressions finite. We therefore introduce
the following Laplace transformed wave functions,
$\Phi^\dagger(\zeta)$ and $\Psi(\eta)$, 
\beq\label{LapPhiPsi}
\Phi^\dagger(\zeta) \ = \
\int\limits_0^\infty \! d L \, e^{-L \zeta} \Psi^\dagger(L)
\, - \, \lambda (\zeta) \, ,
\qquad
\Psi(\eta) \ = \
\int\limits_0^\infty \! d L \, e^{-L \eta} \Psi(L) \, .
\eeq
The commutation relation of the wave functions are 
\beq\label{CommutePhiPsi}
[ \, \Psi (\eta) \, , \, \Phi^\dagger (\zeta) \, ]  \ = \ 
\delta ( \eta , \zeta ) ,
\qquad
[ \, \Phi^\dagger (\zeta_1) \, , \, \Phi^\dagger (\zeta_2) \, ] \ = \ 
[ \, \Psi (\eta_1) \, , \, \Psi (\eta_2) \, ] \ = \ 0 , 
\eeq
where $\delta(\eta,\zeta) = 1/(\eta+\zeta)$
is the Laplace transformation of $\delta(L-L')$.
Then, the universal part of the disk amplitude is given by
\beq\label{DiskAmp3}
F_1^{{\rm univ}(0)} (\zeta;\cc) \ = \
\lim_{\T \rightarrow \infty}
\vac e^{- \T \Hdisk} \Phi^\dagger (\zeta) \cuum .
\eeq

\begin{figure}
  \centerline{\hbox{\psfig{figure=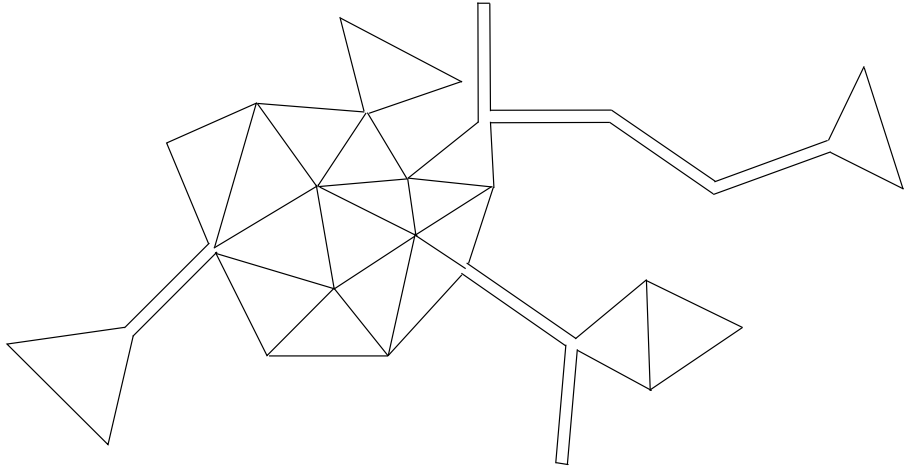,width=12.0cm}}}
  \caption[fig2]{A ``typical'' triangulation with the topology of the 
    ``disk''}
  \label{fig2}
\end{figure}

The actual form of the Hamiltonian can be derived from 
a careful study of a set of Schwinger-Dyson equations, also
called the loop equations, in the context of dynamical
triangulations \cite{watabiki1}.  In Fig.\ \ref{fig2} we
have shown a ``typical'' triangulation with a closed boundary and 
the topology of the disk. This is a concrete realization of the 
surface shown in Fig.\ \ref{fig1}.  In the next subsection we will 
use such triangulations, with more general topology, to derive 
a string field Hamiltonian.  By this procedure every step
is well defined and the lattice spacing $\ep$ can be taken to
zero in an unambiguous way. 

However, it is possible to determine the
continuum Hamiltonian from the following simple
observations: one form of the loop equations (again derived 
from dynamical triangulations or matrix models), where the continuum
limit is already taken is the following
\beq\label{SDf}
0 \ = \
- \, \omega (\zeta,\cc)
\, + \, \frac{\prt}{\prt \zeta} ( F_1^{{\rm univ}(0)} (\zeta;\cc) )^2 ,
\eeq
where
\beq\label{Omega}
\omega (\zeta,\cc)  \  = \  3 \zeta^2 \, - \, \frac{3}{4} \cc .
\eeq
On the other hand, we can derive a kind of Schwinger-Dyson equation
from Hamiltonian formalism if we assume that the limit \rf{DiskAmp3}
is smooth. Under this assumption we have
\beq\label{SD}
0 \ = \  - \, \lim_{\T \rightarrow \infty} \frac{\prt}{\prt \T}
\vac e^{- \T \Hdisk} \Phi^\dagger (\zeta) \cuum
\ = \  \lim_{\T \rightarrow \infty}
\vac e^{- \T \Hdisk} \Hdisk \Phi^\dagger (\zeta) \cuum \, .
\eeq
Comparing \rf{SDf} and \rf{SD} and using \rf{DiskAmp3}, one can expect
\beq\label{Heq}
[ \, \Hdisk \, , \, \Phi^\dagger (\zeta) \, ]  \  =  \  
- \, \omega (\zeta,\cc)
\, + \, \frac{\prt}{\prt \zeta} ( \Phi^\dagger (\zeta) )^2 , 
\eeq
if \rf{VacuumCondition} is satisfied.
Eq.\ \rf{Heq} with the condition \rf{VacuumCondition} leads to
the well-defined Hamiltonian \cite{watabiki1},
\beq\label{Hdisk}
\Hdisk \  = \
\intdzeta \Bigl\{
\, -  \, \omega (\zeta,\cc) \Psi (-\zeta)
\, -  \, ( \Phi^\dagger (\zeta) )^2
\frac{\prt}{\prt\zeta} \Psi (-\zeta) \Bigr\} \, ,
\eeq
where we assume that $\Psi(\eta) \cuum = 0$.
In the expression \rf{Hdisk} regularization is unnecessary since
any number of commutators of $\Hdisk$ with $\Phi^\dg(\zeta)$ or $\Psi(\zeta)$
is finite due to the subtraction of $\lambda(\zeta)$ in \rf{LapPhiPsi}.

\vspace{12pt}

After this brief review of known results we turn to the string
mode expansion. The universal part of the disk amplitude,
$F_1^{\rm univ(0)} (\zeta;\cc)$, has the following expansion
around $\zeta = 0$, 
\beq\label{DiskAmp4}
F_1^{\rm univ(0)} (\zeta;\cc) \ = \
\zeta^{3/2} - \frac{3}{8} \cc \zeta^{-1/2}
+ \sum_{l=1,3,5,...} \zeta^{-l/2-1} f_1^{(0)} (l;\cc) \, , 
\eeq
where $f_1^{(0)}(l;\m)$ are known numbers.
Therefore, we may assume that
the string creation operators $\Phi^\dagger(\zeta)$ and $\Psi(\zeta)$
can be expanded as
\beq\label{StringMode1}
\Phi^\dagger (\zeta) \  =  \
\sum_{l=-5}^\infty \zeta^{-l/2-1} \phi^\dagger_l \, ,~~~~~
\Psi(-\zeta)= \sum_{l=1}^\infty \zeta^{l/2} \phi_l \, ,
\eeq
where the creation and annihilation operators $\phi^\dg_l$ and
$\phi_l$, $l \geq 1$, satisfy the commutation relation,
\beq\label{Commutepsi}
[ \, \phi_l \, , \, \phi^\dagger_{l'} \, ]  \ = \ \delta_{l,l'} \, ,
\qquad
[ \, \phi^\dagger_l \, , \, \phi^\dagger_{l'} \, ] \ = \
[ \, \phi_l \, , \, \phi_{l'} \, ] \ = \ 0 \, .
\eeq
These commutation relations reproduce \rf{CommutePhiPsi}.
Substituting \rf{StringMode1} into \rf{Heq} we find
\beq\label{StringMode2}
\phi^\dagger_{-5} = 1 , ~~~
\phi^\dagger_{-1} = - \frac{3}{8} \cc , ~~~
\phi^\dagger_{-4} = \phi^\dagger_{-3} = \phi^\dagger_{-2} =
\phi^\dagger_0 = 0 \, .
\eeq
Somewhat surprising {\it it is not consistent to have} $\phi_{2n}^\dg =0$, as
one would naively expect from the mode expansion \rf{DiskAmp4}
of the disk amplitude. Such a choice would force $\phi_{2n+1}^\dg$ to
be constants, in contradiction with the fact that $\Phi^\dg(\zeta)$ is
a non-trivial operator.
Thus, we have
\beq\label{StringMode3}
\Phi^\dagger (\zeta)
\ = \ \zeta^{3/2} - \frac{3}{8} \cc \zeta^{-1/2}
\, + \, \sum_{l=1}^\infty \zeta^{-l/2-1} \phi^\dagger_l \, .
\eeq
Note that the operators $\phi_l^\dg$, $ l < 1$, which were introduced
asymmetrically in \rf{StringMode1}, have to be constants, not genuine
non-trivial operators.

The analogy of \rf{DiskAmp3} is now
\beq\label{DiskAmp5}
f_1^{(0)} (l;\cc) \ = \
\lim_{\T \rightarrow \infty}
\vac e^{- \T \Hdisk} \phi^\dagger_l \cuum \, .
\eeq
Since \rf{DiskAmp5} should satisfy \rf{DiskAmp4},
$f_1^{(0)} (l;\cc) = 0$ for $l = \hbox{even integer}$,
i.e. {\it $\phi^\dagger_l$ with even $l$ should be considered as a kind of
  null field} in the sense that  any amplitude of the kind
\rf{DiskAmp5} where $\phi_l^\dg$ is replaced by polynomials of 
$\phi^\dagger_{l_i}$'s will vanish in the $T \to \infty$ limit if 
only one $l_i$ is even.
Substituting \rf{StringMode3} into \rf{Heq}, we find
\beq\label{HeqMode}
[ \, \Hdisk \, , \, \phi^\dagger_l \, ] \  =  \
- \, l \, ( \, \frac{9}{128} \cc^2 \delta_{l,2}
\, + \, \phi^\dagger_{l+1}
\, - \, \frac{3}{8} \theta_{l,4} \cc \phi^\dagger_{l-3}
\, + \, \frac{1}{2} \theta_{l,6} \sum_{k=1}^{l-5}
\phi^\dagger_k \phi^\dagger_{l-k-4} \, ) \, ,
\eeq
where $\theta_{l,k}=1$ for $l \ge k$ and $\theta_{l,k}=0$ for $l < k$.
Finally, we obtain from \rf{HeqMode} and \rf{VacuumCondition}, 
\beq\label{HdiskMode}
\Hdisk \ = \
-  \, \frac{9}{64} \cc^2 \phi_2
\, -  \, \sum_{l=1}^\infty \phi^\dagger_{l+1} l \phi_l
\, +  \, \frac{3}{8} \cc \sum_{l=4}^\infty \phi^\dagger_{l-3} l \phi_l
\, -  \, \frac{1}{2} \sum_{l=6}^\infty \sum_{k=1}^{l-5}
\phi^\dagger_k \phi^\dagger_{l-k-4} l \phi_l \, ,
\eeq
where we have used $\phi_l \cuum = 0$ for $l > 0$.
The first term in \rf{HdiskMode} allows
one string to vanishes into the vacuum,
the second and the third terms are  the ``kinetic'' part of $\Hdisk$,
and the fourth term describes the splitting of one string into two strings.

\subsection{String field theory for general topologies}

\begin{figure}
  \centerline{\hbox{\psfig{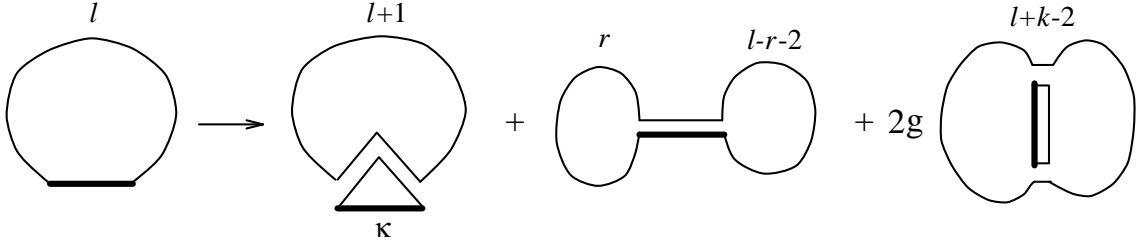}}}
  \caption[fig3]{The possible deformations one can perform starting with 
    a boundary link (shown as a thick line): 
    One can remove a triangle if the link belongs to a triangle, 
    or one can remove a double link, 
    if the link is a part of such double link. 
    The double link can occur in two situations shown symbolically 
    on the figure: 
    if one removes the double link, 
    one boundary will be separated into two boundaries, or 
    two boundaries will be merged into one boundary.} 
\label{fig3}
\end{figure}

Let us generalize the mode expression \rf{HdiskMode} for $\Hdisk$
to the Hamiltonian which produces general amplitudes for orientable surfaces.
This Hamiltonian $\tcH$ has to contain an additional term which
allows the merging of two strings into one. Further, we still
want to maintain the stability of the vacuum, i.e.
\beq\label{VacuumConditionGen}
\tcH \cuum  \ = \  0 .
\eeq
Let us shortly outline how we derive an expression for $\tcH$.
As in the case for $\cH_{\rm disk}$, the starting point is
the special set of Schwinger-Dyson equations called the loop equations.
Whatever Hamiltonian we derive, we want it to reproduce these equations
via \rf{SD}, only with $\tcH$ replacing $\cH_{\rm disk}$.
The loop equations have the graphical representation shown in Fig.\
\ref{fig3}. 
 This figure expresses the possible change at the
discretized level, using dynamical triangulations, when
one deforms the boundary loop $l$. 
If we at the discretized level represent 
the boundary loop as $\Psi^\dg (l)$, and denote 
by $\delta \Psi^\dagger (l)$ the change under a deformation of 
the boundary,  
the figure transforms into the following algebraic equation:
\bea
\langle \delta\Psi^\dagger (l) \prod_i \Psi^\dagger (k_i) \rangle &=&
 \sum_{r=0}^{l-2} \Bigl(
         \langle \Psi^\dagger (r) \Psi^\dagger (l-r-2)
         \prod_i \Psi^\dagger (k_i) \rangle \nn
&&\phantom{\sum_{r=0}^{l-2} \Bigl(}
  + \, \sum_{S} \langle \Psi^\dagger (r)
         \prod_{i \in S} \Psi^\dagger (k_i) \rangle
         \langle \Psi^\dagger (l-r-2)
         \prod_{j \in \bar S} \Psi^\dagger (k_j) \rangle \Bigr)
\nn
&&+ \, 2g \sum_j k_j \langle \Psi^\dagger (l+k_j-2)
         \prod_{i \neq j} \Psi^\dagger (k_i) \rangle
\nn
&&+ \,\k \langle \Psi^\dagger (l+1)
         \prod_i \Psi^\dagger (k_i) \rangle \,
    - \, \langle \Psi^\dagger (l)
         \prod_i \Psi^\dagger (k_i) \rangle \, ,
\label{GravDeform1}
\eea
for $l \ge 1$. In this formula $S$ and $\bar{S}$ are partitions
of the set of indices over which $i$ runs, while $g$
denotes the discretized version of the string coupling
constant and $\kappa$ is related to the cosmological constant
as is clear from Fig.\ \ref{fig3}. In addition we have introduced
the shorthand notation, 
\beq\label{notation}
\langle ~~\equiv ~~\vac \, e^{- T \tcH} \, ~~~~{\rm and}~~~~
\rangle ~~\equiv ~~\cuum \, .
\eeq
We here define 
$\Psi^\dagger (l=0) = 1$ 
in order to make the form of the algebraic equation \rf{GravDeform1} simple. 
Then, we get 
$\delta\Psi^\dagger (l=0) = 0$. 
By taking the discrete Laplace transformation of \rf{GravDeform1}, 
we obtain 
\bea
\langle \del \Psi^\dg (x) \prod_i \Psi^\dagger (z_i) \rangle &=& 
 x^2 \langle \Psi^\dagger (x) \Psi^\dagger (x)
        \prod_i \Psi^\dagger (z_i) \rangle
\nn  
&& + \, x^2\sum_{S} \langle \Psi^\dagger (x)
         \prod_{i \in S} \Psi^\dagger (z_i) \rangle
         \langle \Psi^\dagger (x)
         \prod_{j \in \bar S} \Psi^\dagger (z_j) \rangle
\nn
&& + \, 2g x^2 \langle \Psi^\dagger (x) \Bigl(-x \frac{\prt}{\prt x}
\Psi(\frac{1}{x}) \Bigr) \prod_{i} \Psi^\dagger (z_i) \rangle 
\label{GravDeform3}\\
&& + \, \frac{\k}{x} \langle (\Psi^\dagger (x) -
        x\frac{\prt}{\prt x}\Psi^{\dagger} (x=0)-1)
        \prod_i \Psi^\dagger (z_i) \rangle \,\nn
&& - \, \langle (\Psi^\dagger (x)-1)
        \prod_i \Psi^\dagger (z_i) \rangle \, . 
\nonumber
\eea
From this formula we conclude that
\bea
\del\Psi^\dg (x)& = &x^2 \Bigl(\Psi^\dg (x)\Bigr)^2 \, 
  + \, 2g \Bigl[x^2 \Psi^\dagger (x) \Bigl(-x \frac{\prt}{\prt x}
       \Psi(\frac{1}{x}) \Bigr) \Bigr]^{(+)} \nn
&&+ \, \frac{\k}{x}  \Bigl(\Psi^\dagger (x) \, 
  - \, x \frac{\prt}{\prt x}\Psi^{\dagger} (x=0)-1\Bigr)
  - \, (\Psi^\dagger (x)-1),
\label{GravDeform4}
\eea
where $[x^n]^{(+)}$ means $x^n$ if $n >0$ and zero if $n \leq 0$.

The Schwinger-Dyson equation expresses that the expectation
value of this deformation has to vanish for $T \to \infty$:
\beq\label{GravDeform2}
\lim_{T \to \infty} 
\langle \delta\Psi^\dagger (l) \prod_i \Psi^\dagger (k_i) \rangle\ = \ 0,
\eeq
and string field theory  emerges if we can find a $\tcH$ such that the
vacuum condition is satisfied and \cite{watabiki1}
\beq\label{SFT}
[\tcH,\Psi^\dg(l)] \ = \ - l \del \Psi^\dg(l) \, .
\eeq
The Laplace transform of \rf{SFT} is
\beq\label{STF1}
[\tcH,\Psi^\dg(x) ] \ = \ -x \frac{\prt}{\prt x} \,\del \Psi^\dg (x) \, ,
\eeq
and it allows us to determine $\tcH$ up to a piece which is uniquely
fixed by the vacuum condition.  After taking the continuum limit of 
this equation, where 
\beq\label{CL}
\k = \k_c e^{-\ep^2 \m}, ~~~ g = \ep^5 G, ~~~ x = x_c e^{-\ep \zeta},
\eeq 
define the continuum cosmological constant $\m$, the continuum 
string coupling constant $G$ and our ``continuum'' Laplace transformation
parameter $\zeta$, 
we find the following continuum Hamiltonian \cite{ik,jr,watabiki1}
\beq\label{Hgen}
  \tcH  \ = \
  \intdzeta \Bigl\{
  \,  -   \omega (\zeta,\cc) \Psi (-\zeta)
  \, - \, ( \Phi^\dagger (\zeta) )^2
          \frac{\prt}{\prt\zeta} \Psi (-\zeta)
  \, - \, G \Phi^\dagger (\zeta)
          ( \frac{\prt}{\prt\zeta} \Psi (-\zeta) )^2 \Bigr\} \, .
\eeq

We can now introduce a generating functional, 
\beq\label{GeneratingFun}
   Z_F^{\rm univ} [J]  \ \define \
   \lim_{\T\rightarrow\infty} \vac \,
   e^{- \T \tcH} \,
   \exp[ \int \! \frac{d\zeta}{2 \pi i} \Phi^\dagger(\zeta) J(-\zeta) ]
   \cuum  \, ,
\eeq
and the loop amplitudes are obtained from
\beq\label{GeneralAmp}
   \sum_{h=0}^\infty G^{h+N-1} \,
   F_N^{{\rm univ}(h)} ( \zeta_1, \ldots, \zeta_N ; \cc )
\ = \ \frac{ \delta^{N} }{
      \delta J (\zeta_1) \, \cdots \, \delta J (\zeta_N) } 
    \ln Z_F^{\rm univ} [J] \Bigg|_{J=0} \, .
\eeq
Here $F_N^{{\rm univ}(h)} ( \zeta_1, \ldots, \zeta_N ; \cc )$
denotes the {\it universal} part of loop 
amplitudes,\footnote{There
is some confusion concerning the use of {\it universal}. 
In some articles
the functions $f^{(h)}_N$ to be defined below are denoted universal,
while the amplitudes $F_N^{{\rm univ}(h)}$ 
are called non-universal even they are independent of the cut off.}
i.e. the part which does not contain any cut off dependence.
The cut off dependence appears only in $F_1^{(0)} (\zeta;\cc)$ 
through the function $\l(\zeta)$ and it has already been subtracted
by the shift $\Psi^\dg \to \Phi^\dg$, as explained above. 

The amplitudes $F_N^{{\rm univ}(h)}( \zeta_1, \ldots, \zeta_N ; \cc ) $
are in principle known from the loop equations coming from the
matrix models and they allow expansions of the form, 
\bea
  F_N^{{\rm univ}(h)} (\zeta_1,\ldots,\zeta_N;\cc) &=&
      \left(\Omega_1(\zeta_1) \del_{N,1} + \Omega_2(\zeta_1,\zeta_2)\del_{N,2}
        \right)\del_{h,0}
\label{GeneralAmps}\\
    &&+ \, \sum_{l_i=1,3,5,...}
           \zeta_1^{-l_1/2-1} \cdots \zeta_N^{-l_N/2-1}
           f_N^{(h)} (l_1,\ldots,l_N;\cc) \, ,
\nonumber
\eea
where
\bea
\Omega_1(\zeta)& =& \zeta^{3/2} \, - \, \frac{3}{8} \cc \zeta^{-1/2} \, ,
\label{Omega1}        \\
\Omega_2(\zeta,\eta)&=&
    \frac{1}{4\sqrt{\zeta\eta}(\sqrt{\zeta}+\sqrt{\eta})^2} \, .
\label{Omega2}
\eea
It is seen that $\Omega_1(\zeta)$ and $\Omega_2(\zeta,\eta)$ are precisely the
parts of the universal loop amplitudes which fall off 
slower than $\zeta^{-3/2}$.

A set of ``Hamiltonian'' Schwinger-Dyson equations is obtained as above
by assuming that the limit $T \to \infty$, which defines
$Z_F^{{\rm univ}}[J]$, is smooth. This implies
\beq\label{SDGeneral}
   0 \ = \  - \,
   \lim_{\T\rightarrow\infty} \frac{\prt}{\prt \T} \vac \,
   e^{- \T \tcH} \,
   \exp[ \int \! \frac{d\zeta}{2 \pi i} \Phi^\dagger(\zeta) J(-\zeta) ] \,
   \cuum \, .
\eeq
The differentiation with respect to $T$ brings down the operator $\tcH$,
and we can move it outside the bra-vector 
as a functional differential operator $\tHj$ in $J$, 
\beq\label{SDGeneral2}
   \tHj \, Z_F^{\rm univ} [J] \ = \  0 \, ,
\eeq
\beq\label{Hgen2}
  \tHj \ = \intdzeta \Bigl\{ \,
    - \, \omega (\zeta,\cc) J(-\zeta)
 \, - \, \Bigl( \frac{\prt}{\prt\zeta} J(-\zeta) \Bigr)
         \Bigl( \frac{\delta}{\delta J(\zeta)} \Bigr)^2
 \, - \, G \Bigl( \frac{\prt}{\prt\zeta} J(-\zeta) \Bigr)^2
         \frac{\delta}{\delta J(\zeta)}
  \, \Bigr\} \, .
\eeq

This procedure can be made systematic by introducing the $\star$
operation as follows
\bea
   && \bigl( A_1 A_2 \cdots A_n \bigr)^\star
      \ = \ A_n^\star \cdots A_2^\star A_1^\star \, ,
\nn
   && \bigl( \Phi^\dagger(\zeta) \bigr)^\star
      \ = \ \frac{\delta}{\delta J(\zeta)} \, ,   \qquad
      \bigl( \Psi(\eta) \bigr)^\star
      \ = \ J(\eta) \, , \nn
   && (\vac)^\star = \cuum,~~~~(\cuum)^\star = \vac.
\label{StarOp}
\eea
Then, $\tHj$ in \rf{Hgen2} is obtained after applying the $\star$ operation
to $\tcH$ in \rf{Hgen}.
The condition \rf{VacuumConditionGen}, 
which expresses the stability of the vacuum, becomes
\beq\label{VacuumConditionGen2}
 \vac \tHj  \ = \  0 ,
\eeq
where $\vac J(\eta) = 0$ is equivalent to $\Psi(\eta)\cuum =0$.

Now, let us consider the string mode expansion.
We have seen in the study of the disk amplitude
that the {\it non-trivial operators} $\phi_l^\dg$ were
connected to the amplitudes $f^{(0)}_1(l;\m)$, which have a 
Laplace transform which behaves as $\sum_{l=1,3,\ldots} 
\zeta^{-l/2-1} f^{(0)}_1(l;\m)$. From \rf{GeneralAmps}-\rf{Omega2}
it follows that the Laplace transforms of $f_N^{(h)}(l_1,\ldots,l_N;\m)$
have the same kind of expansion.
Hence, it is natural to introduce a special generating function 
for  the loop-functions  $f_N^{(h)}(l_1,\ldots,l_N;\m)$  
and to expect that they 
have an representation analogous to \rf{DiskAmp5} in terms
of the fields $\phi_l$ and a Hamiltonian $\cH$.
In order to find $\cH$ we first transform $\Omega_1(\zeta)$ and 
$\Omega_2(\zeta,\eta)$ away by writing
\beq\label{univers}
Z_F^{\rm univ} [J] = Z_\Omega [J] Z_f[j],
\eeq
where
\beq\label{Omega3}
\ln Z_\Omega [J]  \  \equiv \  \Omega[J] \, = \, 
\int\! \frac{d\zeta}{2 \pi i} \Omega_1 (\zeta) J(-\zeta) + \frac{G}{2}
\int\! \frac{d\zeta}{2 \pi i} \,\frac{d\eta}{2 \pi i} 
       \Omega_2(\zeta,\eta) J(-\zeta) J(-\eta).
\eeq
With the definition we can write
\beq\label{GeneralAmp2}
   \sum_{h=0}^\infty G^{h+N-1} \,
   f_{N}^{(h)} ( l_1, \ldots, l_N ; \cc )
\ = \ \frac{ \delta^{N} }{
      \delta j_{l_1} \, \cdots \, \delta j_{l_N} } \,
    \ln Z_f [j] \Big|_{j=0} \, ,
\eeq
where we have introduced the notation
\beq\label{modej}
  \frac{\delta}{\delta J(\zeta)} \  =  \
  \Omega_1 (\zeta) + 
  \sum_{l=1}^\infty \zeta^{-l/2-1} \frac{\prt}{\prt j_l} \, ,~~~~~
  J(-\zeta) = \sum_{l=1}^\infty \zeta^{l/2} j_l \, .
\eeq
The Hamiltonian $\tHj$ acting on $Z_F^{\rm univ}[J]$
is naturally related to a Hamiltonian $\Hj$ acting on $Z_f[j]$ by
\beq\label{newham}
\tHj Z_F^{\rm univ} [J] = 0 ~~~~\Rightarrow ~~~~\Hj Z_f[j]=0,
\eeq
where
\beq\label{newhamx}
\Hj = e^{-\Omega[J]} \tHj e^{\Omega[J]} 
    = \tHj -[\Omega[J],\tHj]+ \oh [\Omega[J],[\Omega[J],\tHj]] \, .
\eeq
The vacuum condition \rf{VacuumConditionGen} and \rf{VacuumConditionGen2} 
are replaced by 
\beq\label{VacuumConditionMode}
 \cH \cuum  \ = \  0 \, ,~~~\hbox{and}~~~
 \vac \Hj  \ = \  0 \, ,
\eeq
where $\phi_l \cuum = 0$ and $\vac j_l = 0$ for any $l>0$. 
After some algebra we obtain
\bea
 \Hj &=&
       - \, \frac{9}{64} \cc^2 j_2
    \, - \, \frac{1}{8} G j_4
    \, + \, \frac{3}{16} G \cc j_1 j_2
    \, - \, \frac{1}{16} G^2 j_1 j_1 j_2
\nn
     &&- \, \sum_{l=1}^\infty l j_l \dj{l+1}
    \, + \, \frac{3}{8} \cc \sum_{l=4}^\infty l j_l \dj{l-3}
\nn
     &&- \, \frac{1}{2} \sum_{l=6}^\infty \sum_{k=1}^{l-5}
            l j_l \dj{k} \dj{l-k-4}
    \, - \, \frac{1}{8} G \sum_{k=1}^\infty \sum_{l=1}^{k+3}
            (k-l+4) j_{k-l+4} l j_l \dj{k}  \, .
\label{HgenMode1}
\eea

If we write the star operation in terms of modes 
$\phi_l^\dg$ and $\phi_l$, 
it has the following realization
\beq\label{StarOp2}
  \bigl( \phi^\dagger_l \bigr)^\star \ = \ \frac{\prt}{\prt j_l} \, ,
  \qquad
  \bigl( \phi_l \bigr)^\star \ = \ j_l \, .
\eeq
It enables us to
write $\Hj$ in terms $\phi_l^\dg$ and $\phi_l$:
\bea
  \cH &=&
         - \, \frac{9}{64} \cc^2 \phi_2
      \, - \, \frac{1}{8} G \phi_4
      \, + \, \frac{3}{16} G \cc \phi_1 \phi_2
      \, - \, \frac{1}{16} G^2 \phi_1 \phi_1 \phi_2
\nn
       &&- \, \sum_{l=1}^\infty \phi^\dagger_{l+1} l \phi_l
      \, + \, \frac{3}{8} \cc \sum_{l=4}^\infty \phi^\dagger_{l-3} l \phi_l
\nn
       &&- \, \frac{1}{2} \sum_{l=6}^\infty \sum_{k=1}^{l-5}
              \phi^\dagger_k \phi^\dagger_{l-k-4} l \phi_l
      \, - \, \frac{1}{8} G \sum_{k=1}^\infty \sum_{l=1}^{k+3}
              \phi^\dagger_k (k-l+4) \phi_{k-l+4} l \phi_l \, .
\label{HgenMode2}
\eea
Note that $\cH$ satisfies $\Hdisk = \cH\big|_{G=0}$.

We can finally verify that $Z_f[j]$ has the
following representation in terms of $\cH$ and the modes $\phi_l^\dg$, 
\beq\label{GeneratingFun2}
   Z_f [j]  \ = \
   \lim_{\T\rightarrow\infty} \vac \,
   e^{- \T \cH} \,
   \exp\bigl( \sum_{l=1}^\infty \phi^\dagger_l j_l \bigr) \,
   \cuum  \, ,
\eeq
and in this way $\cH$ {\it is} the Hamiltonian
for the amplitudes $f^{(h)}_N$ in the same
way as $\tcH$ is the Hamiltonian for the amplitudes $F^{{\rm univ}(h)}_N$.
In particular, we can derive $\Hj Z_f[j] =0$, i.e.
the last equation in \rf{newham} from the analogy to \rf{SDGeneral}:
\beq\label{SDGeneral5}
   0 \ = \  - \,
   \lim_{\T\rightarrow\infty} \frac{\prt}{\prt \T} \vac \,
   e^{- \T \cH} \,
   \exp\bigl( \sum_{l=1}^\infty \phi^\dagger_l j_l \bigr) \,
   \cuum \, .
\eeq

\subsection{Relation to $W$-algebras and $\tau$-functions}

The purpose of this subsection is to make the algebraic structure
of $\Hj$ more transparent. Recall the following formal representation
of the so-called $W$-generators: Let the ``current'' $\tJ(z)$
be defined by
\beq\label{current}
\tJ (z) = \sum_{n \in Z} \a_n z^{-n-1} \, ,
\eeq
where
\beq\label{Alpha1}
  \alpha_n \ = \ \left\{
                         \begin{array}{ll}
                            -n x_{-n}             & \mbox{if $n<0$} \\
                            0                     & \mbox{if $n=0$} \\
                            \frac{\prt}{\prt x_n} & \mbox{if $n>0$}
                         \end{array}
                 \right. \, ,  \qquad
[\a_m,\a_n] = m\delta_{m+n,0} \, .
\eeq
The $W$-generators are defined by normal ordering with 
respect to the $\a_n$ generators by the formula, 
\bea 
W^{(k)} (z)& =& \sum_{l=0}^{k-1} \frac{(-1)^l}{(k-l) l!}
\frac{([k-1]_l)^2}{[2k-2]_l} \; \Bigl(\frac{\prt}{\prt z}\Bigr)^l \,
P^{(k-l)} (z) \, ,
\label{wdef} \\
P^{(k)}(z) &=& : \Bigl( \frac{\prt}{\prt z} + \tJ (z) \Bigr)^k : \, 1 \, ,
\label{jdef}
\eea
where $[k]_l = k(k-1)\cdots (k-l+1)$. 
The first few $W^{(k)}(z)$'s are
\bea
W^{(1)} (z) &=& \tJ(z) \, , \nn
W^{(2)} (z) &=& \oh : \tJ(z)^2 : \, , \nn
W^{(3)} (z) &=& \frac{1}{3} : \tJ(z)^3 : \, , \label{Wgen}\\
W^{(4)} (z) &=& \oq :\left\{ \tJ(z)^4 + \frac{2}{5} \tJ(z) \prt^2 \tJ(z) -
                \frac{3}{5} (\prt \tJ(z))^2 \right\}: \, , \nonumber
\eea
and each of the $W^{(k)}(z)$'s has a mode expansion, 
\beq\label{Wmode}
W^{(k)} (z) = \sum_{n \in Z} W^{(k)}_n z^{-n-k},
\eeq
i.e. for the first few values of $k$, 
\bea
W^{(1)}_n &=& \a_n \, , \nn
W^{(2)}_n & =& \oh \sum_{a+b=n} : \a_a\a_b: \, ,\nn
W^{(3)}_n & =& \frac{1}{3} \sum_{a+b+c=n} : \a_a\a_b\a_c : \, , \nn
W^{(4)}_n & =& \oq \sum_{a+b+c+d=n} :\a_a\a_b\a_c\a_d: \nn
            &&- \, \oq \, \sum_{a+b=n}\{(a+1)(b+1) 
              - \frac{1}{5} (n+2)(n+3)\} : \a_a\a_b : \, .
\eea
Finally, the so-called $p$-reduced $W$-generators are obtained by
only considering modes of $W^{(k)}$, $W^{(k-2)},....$ with mode numbers which
are multiples of $p$:
\bea
\bW^{(1)}_n &=& 
W^{(1)}_{pn} \, , \nn
\bW^{(2)}_n &=& 
\frac{1}{p} \left\{ W^{(2)}_{pn} + \frac{1}{24} (p^2-1) \delta_{n,0} 
            \right\} \, , \nn
\bW^{(3)}_n &=& 
\frac{1}{p^2} \left\{ W^{(3)}_{pn} + \frac{1}{12} (p^2-1) W^{(1)}_{pn}
              \right\} \, , \nn
\bW^{(4)}_n &=& 
\frac{1}{p^3} \left\{ W^{(4)}_{pn} + \frac{7}{20} (p^2-1) W^{(2)}_{pn} 
                      + \frac{7}{960} (p^2-1)^2 \delta_{n,0}
              \right\} \, . \label{PRWG}
\eea
In case $p=2$, 
expressed in terms of the $\a_n$'s, 
we get for the {\it two-reduced} operators, 
\bea
\bW^{(1)}_{n}&=& \a_{2n} \, , \nn
\bW^{(2)}_n &=& \oh \biggl( \, \oh \sum_{a+b=2n} :\a_a \a_b : 
                       \, + \, \frac{1}{8} \delta_{n,0} \, \biggr) \, , \nn
\bW_n^{(3)} &=& 
                \oq \biggl( \, \frac{1}{3} \sum_{a+b+c=2n} : \a_a\a_b\a_c : 
                       \, + \, \oq \a_{2n} \, \biggr) \, .
\label{notation1}
\eea
We here also introduce the $W$-generators which do not depend on
$\bJ_k$'s (or $\a_{pk}$), i.e., 
\bea
{\bW^{(2)}_n}' &=& \frac{1}{p} \Biggl\{ \, 
    \oh \sum_{a \atop (p n)} \sum_{l \atop (0 \bmod p)}\! 
      : \a^{[l_1]}_{a_1} \a^{[l_2]}_{a_2} : \, 
  + \, \frac{1}{24} (p^2-1) \delta_{n,0} \Biggr\} \, , \nn
{\bW^{(3)}_n}' &=& \frac{1}{p^2} \Biggl\{ \, 
    \frac{1}{3} \sum_{a \atop (p n)} \sum_{l \atop (0 \bmod p)}\! 
      : \a^{[l_1]}_{a_1} \a^{[l_2]}_{a_2} \a^{[l_3]}_{a_3} : 
  \Biggr\} \, , \nn
{\bW^{(4)}_n}' &=& \frac{1}{p^3} \Biggl\{ \, 
    \oq \sum_{a \atop (p n)} \sum_{l \atop (0 \bmod p)}\! 
      : \a^{[l_1]}_{a_1} \a^{[l_2]}_{a_2} \a^{[l_3]}_{a_3} \a^{[l_4]}_{a_4} :
\, - \, \frac{p}{8} : \biggl[ 
    \sum_{a \atop (p n)} \sum_{l \atop (0 \bmod p)}\! 
      \a^{[l_1]}_{a_1} \a^{[l_2]}_{a_2} \biggr]^2 : \nn
&&\phantom{\frac{1}{p^3} \Biggl\{ \, }
  - \oq \sum_{a \atop (p n)} \sum_{l \atop (0 \bmod p)}\! 
    \biggl( \frac{1}{12} (p-6)(p^2-1) + l_1 l_2 \biggr)
      : \a^{[l_1]}_{a_1} \a^{[l_2]}_{a_2} : \nn
&&\phantom{\frac{1}{p^3} \Biggl\{ \, }
  - \frac{1}{5760} (p^2-1)(p-2)(p-3)(5p+7) \delta_{n,0}
  \Biggr\} \, , \label{PRWG2}
\eea
where we have introduced the notation, 
\bea
  \sum_{a \atop (p n)} &\define& 
  \sum_{a_1} \cdots \sum_{a_k} 
~~~\hbox{with}~~\sum_{i=1}^k a_i = p n \, ,
\nn
  \sum_{l \atop (0 \bmod p)} &\define& 
  \sum_{l_1=1}^{p-1} \cdots \sum_{l_k=1}^{p-1}
~~~\hbox{with}~~\sum_{i=1}^k l_i = 0 \bmod p \, ,
\eea
and 
\beq\label{alphanot}
\a^{[i]}_n =
\left\{ \begin{array}{ll}
                  \a_n   & (n = i \bmod p) \\
                  0      & (\mbox{otherwise})
               \end{array}
       \right. \, .
\eeq

We now give the relation 
between the current $\tJ(z)$ and the string modes. 
We introduce the identification, 
\beq\label{Alpha}
  nx_n \ = \  \sqrt{G}\,n j_n \, + \, \frac{v_n}{\sqrt{G}} \, ,~~~~~
  \frac{\prt}{ \prt x_n} = \frac{1}{\sqrt{G}} \; \frac{\prt}{\prt j_n} \, .
\eeq
Here $v_n$ denotes a possible shift 
and depends on the cosmological constant $\cc$. 
Then we can write
\beq\label{jz}
  \a(z) \ = \ 
  \frac{1}{\sqrt{G}} \Bigl( v(z) + \frac{\delta}{\delta j(z)} \Bigr)
  \, + \, \sqrt{G} \frac{\prt}{\prt z} \Bigl( z j(z) \Bigr) \, ,
\eeq
and
\beq\label{jstar}
  {\a}^\star(z) \ = \
  \frac{1}{\sqrt{G}} \Bigl( v(z) + \phi^\dagger(z) \Bigr)
  \, + \, \sqrt{G} \frac{\prt}{\prt z} \Bigl( z \phi(z) \Bigr) \, ,
\eeq
where 
\bea
&&
\frac{\delta}{\delta j(z)} \ = \ 
\sum_{n=1}^\infty \frac{\prt}{\prt j_n} z^{-n-1} \, , ~~~
j(z) \ = \  
\sum_{n=1}^\infty j_n z^{n-1} \, ,
\nn
&&
\phi^\dagger(z) \ = \  \sum_{n=1}^\infty \phi_n^\dagger z^{-n-1} \, , ~~~
\phi(z) \ = \  \sum_{n=1}^\infty \phi_n z^{n-1} \, , ~~~
v(z) \ = \ \sum_{n=1}^\infty v_n z^{n-1} \, . 
\label{phijz}
\eea
The mode expansion of 
the string field and its current, $\Phi^\dg(\zeta)$ and $J(\zeta)$, 
are expressed by 
the modes $\phi^\dagger(z)$ and $j(z)$ 
if we identify $\zeta = z^p$, i.e. we write 
\bea
&&
\frac{\delta}{\delta J(\zeta)} \ = \  
\frac{1}{z^{p+k-2}} \Bigl( v(z) + \frac{\delta}{\delta j(z)} \Bigr) \, , ~~~
J (-\zeta) \ = \  
z^k j (z) \, ,
\nn
&&
\Phi^\dagger (\zeta) \ = \  
\frac{1}{z^{p+k-2}} \Bigl( v(z) + \phi^\dagger (z) \Bigr) \, , ~~~
\Psi (-\zeta) \ = \  
z^k \phi (z) \, ,
\label{PhiJzeta}
\eea
where $k$ is integer 
and will be uniquely determined for each $(p,q)$ model. 

In the present case of pure gravity we have
$k=1$, and 
\beq\label{valu}
v_n \ = \ - \, \frac{3\cc}{8} \delta_{n,1} \, + \, \delta_{n,5} \, .
\eeq
With these definitions the Hamiltonian can be rewritten
in terms the two-reduced operators as
\beq\label{HgenModeW}
  \Hj \ = \  - \,2 \sqrt{G} \, \bW_{-2}^{(3)} \,
             + \, \frac{1}{2\sqrt{G}} \a_6 \,
             - \, \frac{3\cc}{8\sqrt{G}} \a_2 \, .
\eeq

The stability of the vacuum is expressed by \rf{VacuumConditionMode}.
In the eq.\ \rf{HgenModeW} the second and the third terms are
necessary and sufficient for the condition \rf{VacuumConditionMode}
to be satisfied.
Recall the observation that $\phi^\dg_l$ was a null field
for $l$ an even integer. By \rf{StarOp2} and \rf{Alpha}
this translates into the statement, 
\beq\label{surprise}
\bJ_n Z_f[j] = 0 ,~~~{\rm or}~~~
\frac{\prt}{\prt j_{2n}} Z_f[j] = 0 ,
\eeq
for $n > 0$.
If we denote the part of $\bW^{(2)}_n$ which does not depend on
$\bJ_k$'s (or $\a_{2k}$), $k >0$ by ${\bW_n^{(2)}}'$, i.e.
\beq\label{ReducedW}               
{\bW_n^{(2)}}' = \  \frac{1}{4} \Bigl(
                     \, \sum_{l+m+1=n} : \a_{2l+1} \a_{2m+1} : \,
                   + \, \frac{1}{4} \delta_{n,0} \, \Bigr) \, ,
\eeq
the Hamiltonian \rf{HgenModeW} can also rewritten as
\bea
  \Hj \ &=& \
  - \, 2\sqrt{G} \,
    \Bigl( \, \frac{1}{12} \sum_{a+b+c=-2} : \bJ_a \bJ_b \bJ_c\!: \,
           + \sum_{a+b=-2} \bJ_a {\bW_b^{(2)}}' \, \Bigr) \nn
&&  + \, \frac{1}{2\sqrt{G}} \bJ_3 \,
  - \, \frac{3\cc}{8\sqrt{G}} \bJ_1 \, , 
\label{HgenModeW2}
\eea
where
since $\Hj Z_f[j] =0$ and $\bJ_n Z_f[j]=0$ for $n >0$ (see \rf{surprise}) 
this implies that ${\bW_n^{(2)}}' Z_f[j]=0$  for $n \geq -1$.
These two conditions are generally believed to be sufficient 
to ensure that $Z_f[j]$
is a $\tau$-function \cite{tau,fkn2}. Let us summarize the results as follows:
If we assume stability of the vacuum, i.e. \rf{VacuumConditionMode},
the solution of the Schwinger-Dyson equation \rf{SDGeneral5} satisfies
\beq\label{TauFun}
   Z_f [j]  \ = \  \tau [j] \, ,~~~~{\rm where}~~
   \left\{ \begin{array}{ll}
              \bJ_n \tau = 0           & \mbox{if $n \ge 1$} \\
              {\bW_n^{(2)}}' \tau = 0  & \mbox{if $n \ge -1$}
           \end{array}
   \right.  \, ,
\eeq
i.e. $Z_f[j]$ is (according to general beliefs) a $\tau$-function, 
as indicated by the notation in eq.\  \rf{TauFun}.

\section{Multicritical one-matrix models}

Type $(2,2m-1)$ conformal field theories coupled to quantum gravity
are described by multicritical matrix models 
and we know the expressions for the functions
$\Omega_1(\zeta_1)$,  $\Omega_2(\zeta_1,\zeta_2)$,
$F^{{\rm univ}(h)}_N(\zeta_1,\ldots,\zeta_N;\cc)$ and
$f^{(h)}_N(\zeta_1,\ldots,\zeta_N;\cc)$,
as well as the non-universal part $\l(\zeta)$.
The transfer matrix has been derived in \cite{gk}, 
while the string field theory expressed by 
$\Phi^\dagger(\zeta)$ and $\Psi(\eta)$
has been developed in \cite{watabiki1}. 
We can repeat the calculations in the last section for the $(2,2m-1)$ models
and for all $m \geq 2$ it is now possible to write
\beq\label{mgetwo}
\Hj \ = \ -2 \sqrt{G} \, \bW^{(3)}_{-2} \, + \, Y \, ,
\eeq
where $Y$ is a sum of terms which all contain some operators $\a_{2n}$ 
to the right and therefore annihilate $Z_f[j]$.
It is  possible to adjust the coefficients of the terms in $Y$
without changing the Hamiltonian Schwinger-Dyson equations and as
in pure gravity we can find a $Y$ term such that $\Hj$ in 
addition  satisfies $\vac \Hj = 0$.
The string field and its current are related with the operators $\a_n$ 
through \rf{phijz} and \rf{PhiJzeta} with $k=2m-3$. 
The Hamiltonian \rf{mgetwo} is equivalent to 
that obtained in \cite{watabiki1} 
up to a replacement of the first few of 
$\phi_1^\dagger$, $\phi_2^\dagger$, $\phi_3^\dagger$, $\ldots$ 
with $f^{(0)}_1(1;\m)$, $f^{(0)}_1(2;\m)$, $f^{(0)}_1(3;\m)$, $\ldots$. 
In the case of $G=0$, 
the Hamiltonians in \rf{mgetwo}, 
in ref.\ \cite{watabiki1}, and in ref.\ \cite{ik3}, 
are all equivalent to each other 
up to some derivatives with respect to $z$ in front of $\phi(z)$. 
We postpone  the problem of understanding  this difference 
to future studies. 

If we define \cite{mss}
\beq\label{MultiXn}
  n x_n \ = \  \sqrt{G} n j_n \,
        + \, \frac{v_n}{\sqrt{G}} \sum_{k=0}^{m-1} \delta_{n,4k+1} \, ,
\eeq
$v_1$, $v_5$, $v_9$, $\ldots$ are determined 
by the Schwinger-Dyson equations
and we get 
\beq\label{MultiHgenModeW}
 Y \ = \  \frac{1}{2\sqrt{G}} \sum_{l=0}^{m-1} \sum_{k=\max(0,1-l)}^{m-1} 
             v_{4l+1} v_{4k+1} \a_{4l+4k-2} \,
     + \, \theta_{m,3} \sum_{l=2}^{m-1} \sum_{k=1}^{2l-2}
             v_{4l+1} \a_{4l-2k-3} \a_{2k} \, .
\eeq
This choice of $v_n$ amounts to a specific choice of so-called 
conformal background \cite{mss}. 

\section{The Ising model coupled to gravity}

\begin{figure}
  \centerline{\hbox{\psfig{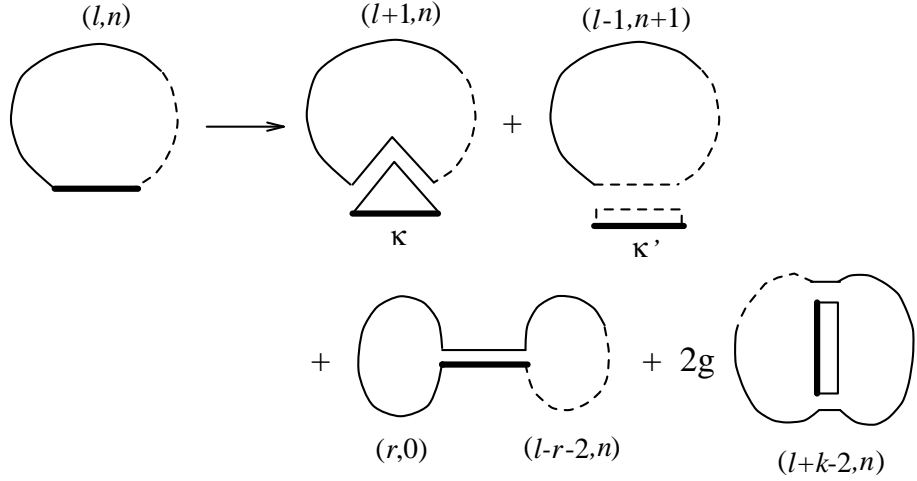}}}
\caption[fig4]{The deformation in the case of Ising spins which 
is almost equivalent to the one for pure gravity. The solid lines 
represent $+$ spins on the boundary, the dashed lines $-$ spins 
on the boundary, while the thick solid line represents the $+$ spin 
which is being deformed. Note that we only consider special spin
configurations on the boundary.}
\label{fig4}
\end{figure}

\begin{figure}
  \centerline{\hbox{\psfig{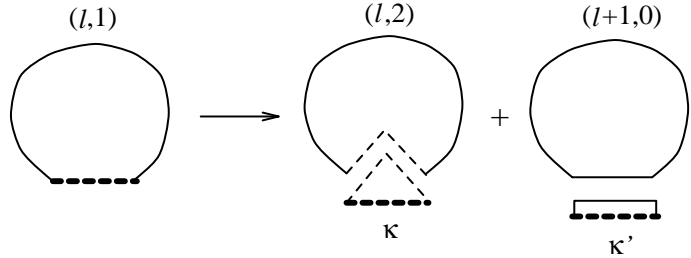}}}
\caption[fig5]{The deformation of a $-$ spin (the thick dashed link), in the 
case where the boundary only has one $-$ spin. The notation is as in Fig.\
\ref{fig4}.}
\label{fig5}
\end{figure}

At the critical point the Ising model coupled to gravity
describes a $(p,q)=(3,4)$ conformal field theory coupled
to gravity. 
A new aspect appears compared to the situation above. 
Let us use the matrix model representation to describe the situation. 
The matrix model action will be  
\beq\label{matrixaction}
S(A_+,A_-) \ = \ N \, {\rm Tr} \,
\Bigl(\, \oh A_+^2 + \oh A_-^2  - \k' A_+ A_- 
         - \frac{\k}{3} A_+^3 - \frac{\k}{3}A_-^3
\,\Bigr) \, , 
\eeq
where $A_+$ and $A_-$ are Hermitian $N\times N$ matrices which
can be associated with ``+'' and ``$-$'' links. 
The matrix model will generate 
dynamical triangulations where the triangles have $+$ or $-$ spin 
and the $A_+ A_-$ term allows us in addition to glue together $+$ and 
$-$ triangles. The boundary configuration is characterized by 
a succession of $+$ and $-$ links.    
It has not yet been possible to solve the discretized theory for
arbitrary boundary spin configurations and we
expect that most such boundary spin configurations are
not important at the critical point. It {\it is} possible to solve the
loop equations for the restricted set of boundary spin configurations
where  a boundary consisting of $n+l$ links has $l$ neighboring $+$ spins
and $n$ neighboring $-$ spins \cite{gn}. Let us denote such a string state
by $\Psi^\dagger_n (l)$. The deformation of
$\Psi^\dagger_n (l)$ at the point of the boundary where
the $+$ and the $-$ links meet  is almost like
the deformation of $\Psi^\dg(l)$ in the case
of pure gravity. The only difference occurs when
the $+$ spin at the boundary is glued to the rest
of the triangulation by means of the interaction term $A_+A_-$
or when we have a single $-$ spin, or vice verse.
We have illustrated these situations in Fig.\ \ref{fig4}
and Fig.\ \ref{fig5}. 
The two equations which replace the string deformation equation of
pure gravity become:
\bea
 \langle \delta\Psi^\dagger_n (l) \prod_i \Psi^\dagger_0 (k_i) \rangle& =&
 \sum_{r=0}^{l-2} \Bigl(
         \langle \Psi^\dagger_0 (r) \Psi^\dagger_n (l-r-2)
         \prod_i \Psi^\dagger_0 (k_i) \rangle \nn
&&\phantom{\sum_{r=0}^{l-2} \Bigl(}
  + \, \sum_{S} \langle \Psi^\dagger_0 (r)
        \prod_{i \in S} \Psi^\dagger_0 (k_i) \rangle
        \langle \Psi^\dagger_n (l-r-2)
        \prod_{j \in \bar S} \Psi^\dagger_0 (k_j) \rangle \Bigr)
\nn
&&+ \, 2g \sum_j k_j \langle \Psi^\dagger_n (l+k_j-2)
         \prod_{i \neq j} \Psi^\dagger_0 (k_i) \rangle \nn
&&+ \, \k \langle \Psi^\dagger_n (l+1)
         \prod_i \Psi^\dagger_0 (k_i) \rangle \, 
  + \, \k' \langle \Psi^\dagger_{n+1} (l-1)
         \prod_i \Psi^\dagger_0 (k_i) \rangle
\nn
&&- \, \langle \Psi^\dagger_n (l)
         \prod_i \Psi^\dagger_0 (k_i) \rangle \, ,
\label{IsingDeform1}
\eea
for $l \geq 1$ and $n \geq 0$, and
\bea
\langle \tilde{\del} \Psi^\dg_1(l) \prod_i \Psi^\dagger_0 (k_i) \rangle& =&
\k \langle \Psi^\dagger_2 (l)
     \prod_i \Psi^\dagger_0 (k_i) \rangle \, 
+ \, \k' \langle \Psi^\dagger_0 (l+1)
     \prod_i \Psi^\dagger_0 (k_i) \rangle 
\nn
&&
- \, \langle \Psi^\dagger_1 (l)
     \prod_i \Psi^\dagger_0 (k_i) \rangle \, ,
\label{IsingDeform2}
\eea
for $l \geq 0$. 
We here define $\Psi^\dagger_0 (l=0) = 1$. 
The notation is as in
\rf{GravDeform1} except that
$\kappa'$ includes in addition the effect of a spin flip, as is clear
from the discussion above and eq.\ \rf{matrixaction}.

Recursively using eqs.\ \rf{IsingDeform1} and \rf{IsingDeform2}, 
it is possible to get a closed equation for $\Psi^\dg_0 (l)$. 
If we introduce a new deformation, 
\bea
\langle \delta^{\hbox{\scriptsize new}}\Psi^\dagger_0 (l)
  \prod_i \Psi^\dagger_0 (k_i) \rangle &\define&
 \sum_{r=1}^{l-1} \Bigl(
         \langle \delta\Psi^\dagger_0 (r) \Psi^\dagger_0 (l-r-1)
         \prod_i \Psi^\dagger_0 (k_i) \rangle \label{IsingDeform3}\\
&&\phantom{\sum_{r=1}^{l-1} \Bigl(}
 + \, \sum_{S} \langle \delta\Psi^\dagger_0 (r)
         \prod_{i \in S} \Psi^\dagger_0 (k_i) \rangle
         \langle \Psi^\dagger_0 (l-r-1)
         \prod_{j \in \bar S} \Psi^\dagger_0 (k_j) \rangle \Bigr)
\nn
&&
 + \, 2g \sum_j k_j \langle \delta\Psi^\dagger_0 (l+k_j-1)
         \prod_{i \neq j} \Psi^\dagger_0 (k_i) \rangle \nn
&&
 + \, \langle \bigl( \frac{\k'}{\k} \delta\Psi^\dagger_0 (l)
                     + \k \delta\Psi^\dagger_0 (l+2)
                     - \delta\Psi^\dagger_0 (l+1)
              \bigr)
         \prod_i \Psi^\dagger_0 (k_i) \rangle \nn
&& 
 - \, \k' \langle \bigl( \delta\Psi^\dagger_1 (l) - 
                  \frac{\kappa'}{\kappa} \tilde{\del}\Psi^\dagger_1 (l-1)
                  \bigr)
         \prod_i \Psi^\dagger_0 (k_i) \rangle \, ,
\nonumber
\eea
then we get 
\beq
\langle \delta^{\hbox{\scriptsize new}}\Psi^\dagger_0 (l)
  \prod_i \Psi^\dagger_0 (k_i) \rangle =
\sum_{
               0 \le r, ~ 0 \le s \atop r+s \le l-3}
         \langle \Psi^\dagger_0 (r) \Psi^\dagger_0 (s)
                 \Psi^\dagger_0 (l-r-s-3)
                 \prod_i \Psi^\dagger_0 (k_i) \rangle \,
    + \, \ldots\ldots
\label{IsingDeform4}
\eeq
In Appendix 1 we present the exact form of (the Laplace transformed of)
$\delta^{\hbox{\scriptsize new}}\Psi^\dagger_0 (l)$. The important thing
to note here is that the terms on the right hand side of eq. 
\rf{IsingDeform4} only contain reference to $\Psi^\dg_0(l)$, i.e.
we can develop a {\it string field theory for $+$ loops only}.
Of course it would be desirable to have a complete string field theory
which includes all possible spin loops, but it seems presently quite
complicated to develop such theory.\footnote{R.\ Nakayama has taken the 
first step in this direction by formulating a string field theory which
involves + loops {\em and} $-$ loops \cite{Rnakayama}.} 
Here we confine ourselves 
to the sector of the complete string field theory which only involves
$+$ loops. As shown above this is possible, but the price we pay is
that the deformation $\delta^{\hbox{\scriptsize new}}\Psi^\dagger_0(l) $
looses the simple relation to geodesic distance which was present in the
case of pure gravity: moving one triangle ``forward''  all around the
boundary should correspond to a step of one lattice unit, i.e. a
step of length one in geodesic (lattice) units.

The two set of Schwinger-Dyson equations which determine the
loop correlations have the form \cite{gn}, 
\beq\label{IsingSD}
\lim_{T \to \infty}\langle \delta\Psi^\dagger_n (l)
\prod_i \Psi^\dagger_0 (k_i) \rangle
\ = \ 0 \, ,~~~~~
\lim_{T \to \infty}\langle \tilde{\delta}\Psi^\dagger_1 (l)
\prod_i \Psi^\dagger_0 (k_i) \rangle
\ = \ 0 \, .
\eeq
In particular, 
\beq\label{IsingSD2}
\lim_{T \to \infty} \langle 
\delta^{\hbox{\scriptsize new}}\Psi^\dagger_0 (l)
\prod_i \Psi^\dagger_0 (k_i) \rangle
\ = \ 0 \, , 
\eeq
will be reproduced in a SFT context if we find a Hamiltonian $\cH$
such that
\beq
[ \cH , \Psi^\dagger_0(l) ] \ = \ 
- \, l \delta^{\hbox{\scriptsize new}}\Psi^\dagger_0(l) \, .
\eeq
We find  after some tedious algebra along the lines discussed
in the last section after taking the continuum limit
\bea
\Hj \ &=&
  - \, \frac{1}{3}     \sum_{a=b+c+d+9} a j_a \dj{b} \dj{c} \dj{d} \,
  - \, \frac{G}{2}    \sum_{a+b=c+d+9} a j_a b j_b \dj{c} \dj{d}
\nn
&&- \, \frac{G^2}{3} \sum_{a+b+c=d+9} a j_a b j_b c j_c \dj{d} \,
  - \, \frac{G}{6} \sum_{a=b+9} (ab+22) a j_a \dj{b}
\nn
&&- \, v_7 \sum_{a=b+c+2} a j_a \dj{b} \dj{c} \,
  - \, v_1 \sum_{a=b+c+8} a j_a \dj{b} \dj{c}
\nn
&&- \, G v_7 \sum_{a+b=c+2} a j_a b j_b \dj{c} \,
  - \, G v_1 \sum_{a+b=c+8} a j_a b j_b \dj{c}
\nn
&&- \, v_7^2 \sum_{a+5=b} a j_a \dj{b} \,
  - \, 2 v_7 v_1 \sum_{a=b+1} a j_a \dj{b} \,
  - \, v_1^2 \sum_{a=b+7} a j_a \dj{b}
\nn
&&- \, 2G^3 \Bigl\{
             ( j_1 + \frac{v_1}{G} )^3 j_6 \,
      + \, 5 ( j_1 + \frac{v_1}{G} )^2 j_2 j_5 \,
      + \, 6 ( j_1 + \frac{v_1}{G} )^2 j_3 j_4
\nn
&&\phantom{ - \, 2G^3 \Bigl\{ }
      + \, 8 ( j_1 + \frac{v_1}{G} ) j_2^2 j_4 \,
      + \, 9 ( j_1 + \frac{v_1}{G} ) j_2 j_3^2 \,
      + \, 4 j_2^3 j_3 \, \Bigr\}
\nn
&&- \, \frac{4G^2}{3} \Bigl\{
           14 ( j_1 + \frac{v_1}{G} ) j_8 \,
      + \, 14 j_2 ( j_7 + \frac{v_7}{7G} ) \,
      + \,  9 j_3 j_6 \, + \, 5 j_4 j_5 \, \Bigr\}
\nn
&&- \, \frac{v_7^2}{G} (  \dj{1} \dj{4} + \dj{2} \dj{3} ) \,
  - \, \frac{v_7^3}{3G} \dj{12} \,
  - \, \frac{v_7^2 v_1}{G} \dj{6} \, + \, 9G Y \, .
\label{IsingH1}
\eea
Here $Y$ denotes a sum of terms which all contain some operators $\dj{3n}$
in the same way as $Y$ in the last section 
denoted a sum of terms which all contains some operators $\dj{2n}$. 
They are singled out because
\beq\label{j3surprise}
\dj{3n} Z_f [j] = 0,
\eeq
the analogy of \rf{surprise} for $p=3$ in the $(p,q)$ system.
The string field and its current are related with the operators $\a_n$ 
through \rf{phijz} and \rf{PhiJzeta} with $k=1$. 

It is now possible to proceed as in the last section and introduce
\beq\label{IsingXn}
  n x_n \ = \  \sqrt{G} n j_n \,
          + \, \frac{v_1}{\sqrt{G}} \delta_{n,1} \,
          + \, \frac{v_7}{\sqrt{G}} \delta_{n,7} \, ,
~~
  \frac{\prt}{\prt x_n} \ = \
  \frac{1}{\sqrt{G}} \frac{\prt}{\prt j_n} \, ,
\eeq
where the values of $v_1$ and $v_7$ are $-\cc/3$ and $1$, respectively 
(but it is not important for the arguments to follow).
We then find that \rf{IsingH1} is rewritten as
\beq\label{IsingH2}
\Hj \ = \  - \, 9G \, \bW^{(4)}_{-3} \, + \, Y \, .
\eeq
Using $\dj{3n} Z_f [j] = 0$, 
the constraint $\Hj Z_f[j] = 0$ becomes
\bea
  \bW^{(4)}_{-3} \, Z_f [j] & = &
  \Bigl( \, \sum_{n=1}^\infty \alpha_{-3n} \bW^{(3)}_{n-3} \,
         - \, \frac{1}{3} \sum_{n=2}^\infty \sum_{a+b=-3n}\!
               :\alpha^{(0)}_a \alpha^{(0)}_b -
                \alpha^{(1)}_a \alpha^{(2)}_b\!: \bW^{(2)}_{n-3} \,
 \Bigr) Z_f [j]\nn
& =  & 0 \, .\label{IsingC1}
\eea
In this way we finally  arrive at the following constraints:
\beq\label{IsingC2}
  \bW_{n-3}^{(3)} \, Z_f [j] \ = \
  \bW^{(2)}_{n-2} \, Z_f [j] \ = \ \bW^{(1)}_{n-1}Z_f [j] \ =\ 0
\qquad \hbox{for $n \ge 1$.}
\eeq

In order to satisfy the vacuum condition \rf{VacuumConditionMode},
we may try to determine $Y$ in \rf{IsingH2} as was
done in the pure gravity case.
However, {\it in Ising case it is impossible to find a $Y$ such
that the vacuum condition \rf{VacuumConditionMode} is satisfied}, 
since we see from the last line in eq.\ \rf{IsingH1} 
that $\bW_{-3}^{(4)}$ contains a term
\beq\label{unwanted}
- \frac{v_7^2}{G} \; \frac{\prt}{\prt j_1} \frac{\prt}{\prt j_4},
\eeq
which has no reference to $\prt/\prt j_{3n}$.
But in the case of the Ising model we have a larger freedom
to add terms while still satisfying the Schwinger-Dyson
equations since $Z_f[j]$ is  not only annihilated by $\bW^{(1)}_n$ operators
but also by $\bW^{(2)}_n$  operators. We can use these to
modify the Hamiltonian without modifying the Schwinger-Dyson equations.
Potential candidate terms which have the right dimension are
properly normal ordered terms like
\beq\label{termz}
\int \frac{dz}{2\pi i}:(\bW^{(2)} (z))^2:
\eeq
and similar products which involve $\bW^{(2)}(z)(\bW^{(1)}(z))^2$,  
etc. However, these last terms  all contain  $\bW^{(1)}(z)$ and 
they can be absorbed in the definition of $Y$. In this way the 
coefficient of the term \rf{termz} is uniquely fixed by the 
requirement that it should 
cancel the term  \rf{unwanted}, and we get 
\beq\label{IsingH3}
\Hj \ = \  - \, 9G X \, + \, Y \, ,~~~~
X \ = \ \bW^{(4)}_{-3} \,
        - \, \sum_{n=2}^\infty \bW^{(2)}_{-n} \bW^{(2)}_{n-3} \, , 
\eeq
where $Y$ is still undetermined. After the
elimination of \rf{unwanted}, the situation for $\Hj$ in \rf{IsingH3}
is the same as in the $p=2$ case and the term $Y$ is now uniquely determined 
by requiring that the vacuum condition \rf{VacuumConditionMode} is satisfied. 
We find
\beq\label{IsingY}
  Y \ = \     \frac{v_7^3}{3G^{1/2}} \a_{12} \,
         + \, \frac{v_7^2 v_1}{G^{1/2}} \a_6 \,
         + \, v_7^2 \a_2 \a_3 \, .
\eeq
Therefore, we obtain
\beq\label{IsingHgenModeW}
  \Hj \ = \
  - \, 9G \, \Bigl( \,
             \bW^{(4)}_{-3} \,
       - \, \sum_{n=2}^\infty \bW^{(2)}_{-n} \bW^{(2)}_{n-3} \, \Bigr) \,
       + \, \frac{v_7^3}{3G^{1/2}} \a_{12} \,
       + \, \frac{v_7^2 v_1}{G^{1/2}} \a_6 \,
       + \, v_7^2 \a_2 \a_3 \, .
\eeq
The explicit form of $\cH$ corresponding to $\Hj$
is given in the Appendix 2.

Since
\bea
  \bW^{(4)}_{-3} \, - \, \sum_{n=2}^\infty \bW^{(2)}_{-n} \bW^{(2)}_{n-3}
 & =& - \, \frac{1}{216} \sum_{a+b+c+d=-3} : \bJ_a \bJ_b \bJ_c \bJ_d \!:
\label{IsingHgenModeW2}\\
&&+ \, \frac{1}{6} \sum_{a+b+c=-3} : \bJ_a \bJ_b \!: \rW{2}_c \,
  + \, \sum_{a+b=-3} \bJ_a \rW{3}_b \, ,
\nonumber
\eea
we conclude that there exists a Hamiltonian with a stable vacuum such that
the Hamiltonian Schwinger-Dyson equations has a solution $Z_f[j]$ which is 
characterized by
\beq\label{IsingTauFun}
   Z_f [j]  \ = \  \tau [j] \, ,
\qquad \left\{ \begin{array}{ll}
                  \bJ_n \tau = 0     & \mbox{if $n \ge 1$} \\
                  \rW{2}_n \tau = 0  & \mbox{if $n \ge -1$} \\
                  \rW{3}_n \tau = 0  & \mbox{if $n \ge -2$}
               \end{array}
       \right.  \, .
\eeq
These conditions are believed to imply that $Z_f[j]$ is a $\tau$-function,
as we have indicated by the notation used in eq.\ \rf{IsingTauFun}.

\section{Gravity coupled to ($p$,$q$) conformal fields}

The procedures outlined  above can be generalized to any $(p,q)$ model
coupled to gravity. 
The algebraic details are rather tedious so
we will confine ourselves to state the results.

We define as before
\beq\label{pqXn}
  n x_n \ = \  \sqrt{G} n j_n \, + \, \frac{v_n}{\sqrt{G}}  \, ,
~~
  \frac{\prt}{\prt x_n} \ = \
  \frac{1}{\sqrt{G}} \frac{\prt}{\prt j_n} \, ,
\eeq
and obtain the Hamiltonian
\beq\label{pqHgenModeW}
  \Hj \ = \
  - \, p^{(p-1)} G^{(p-1)/2} X \, + \, Y \, .
\eeq
$Y$ denotes a sum of terms which all contain some operators 
$\frac{\prt}{\prt j_{pn}}$
and will be determined  by the vacuum condition
\rf{VacuumConditionMode}. The problem of a stable vacuum is
as in the Ising model, only more severe. 
Again we find that it can be
repaired by modifying the Hamiltonian with terms which do not
interfere with the requirement that $Z_f[j]$ is a $\tau$-function.
After adding such terms we can find a $Y$
such that $\Hj$ has a stable vacuum.

In the case of  $(p,q)=(4,5)$ we find:
\bea
&&X \ = \    \bW^{(5)}_{-4} \,
         - \, \sum_{n=2}^\infty \bW^{(2)}_{-n} \bW^{(3)}_{n-4} \,
         - \, \sum_{n=3}^\infty \bW^{(3)}_{-n} \bW^{(2)}_{n-4} \, , 
\nn
&&Y \ = \     \frac{v_9^4}{4G^{1/2}} \a_{20} \,
         + \, \frac{v_9^3 v_1}{G^{1/2}} \a_{12} \,
         + \, \frac{3v_9^2 v_1^2}{2G^{1/2}} \a_{4} \,
         + \, v_9^3 ( \a_3 \a_8 + \a_7 \a_4 ) \, , 
\eea
and we can alternative write $X$ as
\bea
X & =&   - \, \frac{7}{4^4 15} \sum_{a+b+c+d+e=-4}
       : \bJ_a \bJ_b \bJ_c \bJ_d \bJ_e : \nn
&&  - \, \frac{1}{48} \sum_{a+b+c+d=-4}
       : \bJ_a \bJ_b \bJ_c : \rW{2}_d \,
\nn
&&   + \, \frac{1}{4} \sum_{a+b+c=-4} : \bJ_a \bJ_b : \rW{3}_c \,
  + \, \sum_{a+b=-4} \bJ_a \rW{4}_b \, ,
\label{pqHgenModeW2}
\eea
which makes it easy to understand that $Z_f[j] = \tau[j]$. 
For the $(5,6)$ model we get
\bea
 &&X \ = \     \bW^{(6)}_{-5} \,
         - \, \sum_{n=2}^\infty \bW^{(2)}_{-n} \bW^{(4)}_{n-5} \,
         - \, \sum_{n=3}^\infty \bW^{(3)}_{-n} \bW^{(3)}_{n-5} \,
         - \, \sum_{n=4}^\infty \bW^{(4)}_{-n} \bW^{(2)}_{n-5}
\label{MinimalX}\\
&&\phantom{X \ = \     }
         +  \, \frac{1}{2} \sum_{n=2}^\infty \sum_{m=2}^\infty
               \Norder \bW^{(2)}_{-n} \bW^{(2)}_{-m} 
                       \bW^{(2)}_{n+m-5} \Norder \,
         +  \, \frac{1}{2} \sum_{n=-1}^\infty \sum_{m=-1}^\infty
               \Norder \bW^{(2)}_{-n-m-5}
               \bW^{(2)}_{n} \bW^{(2)}_{m} \Norder \, ,
\nn
&&Y \ = \     \frac{v_{11}^5}{5G^{1/2}} \a_{30} \,
         + \, \frac{v_{11}^4 v_1}{G^{1/2}} \a_{20} \,
         + \, \frac{2v_{11}^3 v_1^2}{G^{1/2}} \a_{10} \,
         + \, v_{11}^4 ( \a_4 \a_{15} + \a_9 \a_{10} + \a_{14} \a_5 )
\nn
&&\phantom{Y \ = \     }
         + \, 4v_{11}^3 v_1 \a_4 \a_5 \,
         - \, G^{1/2} v_{11}^3 \a_1 \a_2 \a_5 \, .
\label{MinimalY}
\eea
The string field and its current are related with the operators $\a_n$ 
through \rf{phijz} and \rf{PhiJzeta} with $k=1$ for any $(m,m+1)$ model. 

It is seen that an interesting algebraic structure is present since
we can write
\bea
  X &=& - \, \oint \! \frac{d z}{2 \pi i} \oint \! \frac{d s}{2 \pi i} \,
        s^{-p-2} \, \Norder \exp [ - \bW (z,s) ] \Norder
\nn
    &=& \bW^{(p+1)}_{-p} \,
        - \, \frac{1}{2} \sum_{k=2}^{p-1} \sum_{n \in \dbl{Z}}
             \Norder \bW^{(k)}_{-n} \bW^{(p-k+1)}_{n-p} \Norder
\nn
    &&
        + \, \frac{1}{3!} \sum_{k \ge 2, ~ l \ge 2 \atop k+l+1 \le p}
                          \sum_{n,m \in \dbl{Z}}
             \Norder \bW^{(k)}_{-n} \bW^{(l)}_{-m}
                     \bW^{(p-k-l+1)}_{n+m-p} \Norder \,
        + \, \ldots\ldots
\label{Xoperator}
\eea
with
\beq\label{W}
  \bW (z,s) \ = \ \sum_{k=2}^\infty \bW^{(k)} (z) s^k  \, ,
\qquad
  \bW^{(k)} (z) \ = \ \sum_{n \in \dbl{Z}} \bW^{(k)}_n z^{-n-k}  \, .
\eeq
With these definitions 
the Hamiltonian \rf{pqHgenModeW} is valid for any $(p,q)$ model.
It should be noted that one in eq.\ \rf{Xoperator} encounters a problem 
defining the normal ordering of 
the product of more than three $\bW$ operators. 
We also note that one could in principle define 
the following more general $X$ operator, 
\bea
  \Norder \exp [ - \bW\!_{[r]} (z,s) ] \Norder \ = \ 
  1 \, - \, \sum_{k=r}^\infty \sum_{n \in \dbl{Z}} 
  X_{[r]n}^{(k)} z^{-n-k} s^k
\label{XoperatorMod}
\eea
with
\beq\label{WMod}
  \bW\!_{[r]} (z,s) \ = \ \sum_{k=r}^\infty \bW^{(k)} (z) s^k  \, ,
\qquad
  \bW^{(k)} (z) \ = \ \sum_{n \in \dbl{Z}} \bW^{(k)}_n z^{-n-k}  \, .
\eeq
It seems that only $X = X_{[2]-p}^{(p+1)}$ 
plays a role for $(p,q)$ models coupled to quantum gravity. 
We further make the surprising observation  that 
$X_{[1]-p}^{(p+1)} = 0$ for $p=2$ and $p=3$ and conjecture 
that it is true for all $p$.

In ref.\ \cite{mss}, 
the authors give the explicit form of the disk amplitudes 
for any $(p,q)$ model. 
Then we have 
\beq\label{DiskExpansionPQ}
F_1^{{\rm univ}(0)} (\zeta;\cc) \ = \ 
\Omega_1(\zeta) \, + \, \const \cc^{1-\gamma} \zeta^{\gamma-1} 
\, + \, \ldots\ldots 
\, ,
\eeq
where 
$\Omega_1(\zeta)$ is a polynomial with respect to $\cc$ and has the form, 
$\Omega_1(\zeta) = \const \zeta^{1-\gamma} + O(\cc)$. 
Here $\gamma$ is the string susceptibility which has the value 
$\gamma = 1-q/p$ for $p<q$ or $\gamma = 1-p/q$ for $p>q$. 
Since $v(z)$ corresponds to $\Omega_1(\zeta)$, 
$v(z)$ is also a polynomial with respect to $\cc$, i.e. 
we get 
\beq\label{pqvev}
v(z) \ = \  \sum_{0 \le i < h/(2p)} v_{h-2ip} \, z^{h-2ip-1} \, ,
\eeq
from $\dim \cc = \dim \zeta^2$ and $\zeta = z^p$, 
where 
$v_h = +1$ or $-1$. 
By using \rf{DiskAmp3} 
one can compare 
the leading term of $\Omega_1(\zeta)$ in \rf{DiskExpansionPQ} 
with that of $v(z)$ in \rf{PhiJzeta}. 
Then we find $h = p+q+k-1$ if $p<q$. 
On the other hand, $h$ becomes non-integer if $p>q$, 
which means that 
we cannot express the Hamiltonian $\cH$ by the string fields. 

We here require that 
the Hamiltonian for disk amplitude, 
$\Hdisk = \cH\big|_{G=0}$, 
has a tadpole term. 
This requirement is quite natural because 
a tadpole is necessary in order to make a disk topology. 
From \rf{pqHgenModeW}, \rf{Xoperator}, \rf{PRWG}, \rf{wdef}, \rf{jdef}, 
\rf{Alpha1}, and \rf{Alpha}, 
the above requirement leads to the fact that 
one of $v_1$, $v_2$, $\ldots$, $v_{p-1}$ is non-zero. 
In order that this fact is consistent with \rf{pqvev}, 
we reach the following most plausible conjecture 
for the general $(p,q)$-SFT with $p<q$: 
$h = p [q/p] + q$, 
which leads to 
$k = p ( [q/p] - 1 ) + 1$. 
This conjecture is satisfied 
in $(2,3)$-SFT, $(2,2m-1)$-SFT's and $(m,m+1)$-SFT's.

\section{String Field Theory for One-string Propagation}

In \cite{transfer} it was shown that important variables 
in quantum gravity which  refer to geodesic distance 
are conveniently obtained by using the transfer matrix,
\beq\label{TM}
  G(n,m;T) \ = \
 \vac \phi_m e^{- \T \Hone} \phi_n^\dagger \cuum \, ,
\eeq
where $\Hone$ is the Hamiltonian,
which expresses the propagation of one-string state,
\beq\label{Hone}
  \Hone \ = \
  \sum_{l=1}^\infty \phi_l^\dagger
  \Biggl[ \frac{\delta}{\delta \phi_l^\dagger} \cH
  \Biggr]_{\phi_l^\dagger \rightarrow f_l , G \rightarrow 0} \, ,
\eeq
where $f_l \equiv f_1^{(0)}(l;\cc)$ is the disk amplitude introduced earlier.

In pure gravity, we find
\beq\label{HonePure}
  \Hone \ = \
  - \, \sum_{l=1}^\infty \phi^\dagger_{l+1} l \phi_l \,
  + \, \frac{3}{8} \cc \sum_{l=4}^\infty \phi^\dagger_{l-3} l \phi_l \,
  - \, \sum_{l=6}^\infty \sum_{k=1}^{l-5}
       f_{l-k-4} \phi^\dagger_k l \phi_l \, .
\eeq
By using $\phi^\dagger(z)$ and $\phi(z)$,
eq.\ \rf{HonePure} becomes
\beq\label{HonePure2}
  \Hone \ = \
  - \, \intz \frac{1}{z^2} \biggl\{
  \Bigl( - \frac{3}{8} \mu + z^4 + f(z) \Bigr)
  \phi^\dagger(z) \prtz \Bigl( z \phi(z) \Bigr) \biggr\} \, ,
\eeq
where $f(z) = \sum_{n=1}^\infty f_n z^{-n-1}$.
In the general case $(p,q)=(2,2m+1)$, we find
\beq\label{HoneP2a}
  \Hone \ = \ 
  - \, \intz \frac{1}{z^2} \biggl\{ 
  \Bigl( v(z) + f(z) \Bigr) 
  \phi^\dagger(z) \prtz \Bigl( z \phi(z) \Bigr) \biggr\} \, .
\eeq
Using \rf{TM} we obtain the following 
differential equation for the transfer matrix, 
\beq\label{Diff2G}
  \frac{\prt}{\prt T} G(x,y;T) \ = \
  - \, \frac{\prt}{\prt x} \biggl\{ \frac{1}{x^2} 
       \Bigl( v(x) + f(x) \Bigr) G(x,y;T) \biggr\}^{(-)} \, ,
\eeq
where $[x^n]^{(-)}$ means $x^n$ if $n < 0$ and zero if $n \geq 0$.
In pure gravity the differential equation \rf{Diff2G} 
is the same equation that was obtained in \cite{transfer}. 
On the other hand, in the multicritical model the eq.\ \rf{Diff2G} 
is slightly different from that in \cite{gk}. 
We postpone the problem of understanding 
this difference to future studies. 

In the gravity coupled to Ising matter,
the Hamiltonian $\Hone$ has the form
\bea
\Hone \ &=&
  - \, \sum_{a=b+c+d+9}
       f_d f_c \phi^\dagger_b a \phi_a \,
  + \, \sum_{n=4}^\infty \sum_{c+d=3n-9 \atop a=b+3n}
       f_d f_b \phi^\dagger_c a \phi_a
  + \, \frac{1}{2} \sum_{n=4}^\infty \sum_{c+d=3n-9 \atop a=b+3n}
       f_d f_c \phi^\dagger_b a \phi_a
\nn
&&- \, v_7 \, \Bigl( \, 
         2 \, \sum_{a=b+c+2} 
         f_c \phi^\dagger_b a \phi_a \,
    - \, \sum_{n=1}^\infty \sum_{a=b+3n} 
         f_{3n-2} \phi^\dagger_b a \phi_a \, 
    - \, \sum_{n=1}^\infty \sum_{a=b+3n} 
         f_b \phi^\dagger_{3n-2} a \phi_a \, 
\nn
&&\phantom{- \, v_7 \, \Bigl( \,}
    - \, \sum_{n=1}^\infty \sum_{a+b=3n}
         f_b \phi^\dagger_a (3n+2) \phi_{3n+2} \, \Bigr)
\nn
&&- \, v_1 \, \Bigl( \, 
         2 \, \sum_{a=b+c+8}
         f_c \phi^\dagger_b a \phi_a \,
    - \, \sum_{n=3}^\infty \sum_{a=b+3n} 
         f_{3n-8} \phi^\dagger_b a \phi_a \, 
    - \, \sum_{n=3}^\infty \sum_{a=b+3n} 
         f_b \phi^\dagger_{3n-8} a \phi_a \, 
\nn
&&\phantom{- \, v_7 \, \Bigl( \,}
    - \, \sum_{n=1}^\infty \sum_{a+b=3n}
         f_b \phi^\dagger_a (3n+8) \phi_{3n+8} \, \Bigr)
\nn
&&- \, v_7^2 \, \Bigl( \, 
         \sum_{a+5=b} \phi^\dagger_b a \phi_a \,
    - \, \sum_{n=1}^\infty \phi^\dagger_{3n+4} (3n-1) \phi_{3n-1} 
  \, \Bigr)
\nn
&&- \, 2 v_7 v_1 \, \Bigl( \, 
         \sum_{a=b+1} \phi^\dagger_b a \phi_a \,
    - \, \sum_{n=1}^\infty \phi^\dagger_{3n-2} (3n-1) \phi_{3n-1} 
  \, \Bigr)
\nn
&&- \, v_1^2 \, \Bigl( \, 
         \sum_{a=b+7} \phi^\dagger_b a \phi_a \,
    - \, \sum_{n=1}^\infty \phi^\dagger_{3n-2} (3n+5) \phi_{3n+5} 
  \, \Bigr) \, .
\label{HoneIsing}
\eea
In this case we cannot express \rf{HoneIsing} in a simple way in terms of
$\phi^\dagger(z)$ and $\phi(z)$. 
Rather, the natural variables seem to be
$\phi^{[i]\dagger}(z)$ and $\phi^{[i]}(z)$. Using these 
variables \rf{HoneIsing} can be written as
by
\bea
\Hone \ = \ 
  \intz \frac{1}{z^6} \hspace{-8pt} &\Biggl[\hspace{-8pt}&
  \Bigl( \oh ( f^{[0]} )^2 -
         f^{[1]} ( v^{[1]} + f^{[2]} ) \Bigr) 
  \phi^{[0]\dagger} \prtz \bigl( z \phi^{[0]} \bigr)
\label{HoneIsing2}
\\
&&
- \oh ( f^{[0]} )^2 
  \Bigl( \phi^{[1]\dagger} \prtz \bigl( z \phi^{[1]} \bigr)
       + \phi^{[2]\dagger} \prtz \bigl( z \phi^{[2]} \bigr) \Bigr)
\nn
&&
- \Bigl( f^{[0]} ( v^{[1]} + f^{[2]} ) + 
         ( f^{[1]} )^2 \Bigr) 
  \Bigl( \phi^{[1]\dagger} \prtz \bigl( z \phi^{[0]} \bigr)
       + \phi^{[0]\dagger} \prtz \bigl( z \phi^{[2]} \bigr) \Bigr)
\nn
&&
- \Bigl( f^{[0]} f^{[1]} + 
         ( v^{[1]} + f^{[2]} )^2 \Bigr) 
  \Bigl( \phi^{[2]\dagger} \prtz \bigl( z \phi^{[0]} \bigr)
       + \phi^{[0]\dagger} \prtz \bigl( z \phi^{[1]} \bigr) \Bigr)
\nn
&&
- 2 f^{[0]} f^{[1]}
    \phi^{[1]\dagger} \prtz \bigl( z \phi^{[2]} \bigr) 
- 2 f^{[0]} ( v^{[1]} + f^{[2]} )
    \phi^{[2]\dagger} \prtz \bigl( z \phi^{[1]} \bigr) \, \Biggr] \, . 
\nonumber
\eea
In the general case $(p,q)=(3,q)$, we find
\beq\label{HoneP3}
\Hone \ = \ 
  - \, \intz \frac{1}{z^6} \sum_{i=0}^2 \sum_{j=0}^2
       M^{[i,j]} \phi^{[j]\dagger} \prtz \bigl( z \phi^{[i]} \bigr) \, ,
\eeq
where
\bea
&&M^{[0,0]} \ = \ 
    - \, \oh ( v^{[0]} + f^{[0]} )^2 
    + ( v^{[2]} + f^{[1]} ) ( v^{[1]} + f^{[2]} ) \, ,
\nn
&&M^{[1,1]} \ = \ M^{[2,2]} \ = \ 
      \oh ( v^{[0]} + f^{[0]} )^2 \, ,
\nn
&&M^{[0,1]} \ = \ M^{[2,0]} \ = \ 
      ( v^{[0]} + f^{[0]} ) ( v^{[1]} + f^{[2]} )
    + ( v^{[2]} + f^{[1]} )^2 \, ,
\nn
&&M^{[0,2]} \ = \ M^{[1,0]} \ = \ 
      ( v^{[0]} + f^{[0]} ) ( v^{[2]} + f^{[1]} )
    + ( v^{[1]} + f^{[2]} )^2 \, ,
\nn
&&M^{[1,2]} \ = \ 
    2 ( v^{[0]} + f^{[0]} ) ( v^{[1]} + f^{[2]} ) \, ,
\nn
&&M^{[2,1]} \ = \ 
    2 ( v^{[0]} + f^{[0]} ) ( v^{[2]} + f^{[1]} ) \, .
\label{HoneP3m}
\eea
The transfer matrix satisfies the differential equations, 
\beq\label{Diff3G}
  \frac{\prt}{\prt T} G^{[i,j]}(x,y;T) \ = \
  - \, \frac{\prt}{\prt x} \biggl\{ \frac{1}{x^6} 
       \sum_{k=0}^2 M^{[i,k]}(x) G^{[k,j]}(x,y;T) \biggr\}^{(-)} \, .
\eeq

\section{Conclusion}

Starting from first principles, i.e. from models regularized by the use 
of dynamical triangulations, we have shown that it is possible 
to define a string field Hamiltonian
with the following properties:  the vacuum is stable and 
the solution $Z_f[j]$ of the Hamiltonian Schwinger-Dyson
equations is a $\tau$-function of the kind expected from matrix model
considerations.

\begin{figure}
  \centerline{\hbox{\psfig{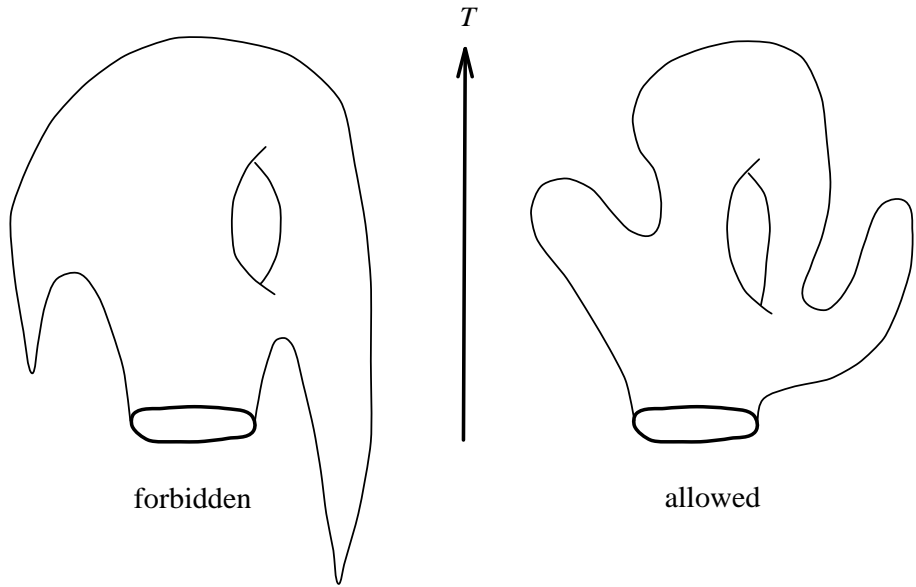}}}
\caption[fig6]{A forbidden configuration and an allowed configuration
for a given choice of proper time slicing indicated by the arrow.}
\label{fig6}
\end{figure}

How unique is this Hamiltonian?
Here we have in the pure gravity case as well as the 
Ising model case relied on an explicit construction of the 
string loop deformation, taken from the formalism of 
dynamical triangulations. Within this formalism, the 
Hamiltonian was unique up to terms which annihilated the 
generating functional $Z_f[j]$.  This ambiguity was 
removed by the requirement of a stable vacuum.
We can formulate the situation as follows: a given 
choice of string loop deformation corresponds to a 
specific choice of proper time slicing of 
our Euclidean space-time and the stability of the 
vacuum rules out the situation that non-trivial 
physics can be created out of the vacuum. 
We have shown such a forbidden situation in Fig.\ \ref{fig6}.

As we have seen the Hamiltonian is also (almost) uniquely determined 
in the case of the $m$th multicritical one matrix models 
since the factor $k=2m-3$. 
However, our transfer matrix is not completely identical 
to that of derived in \cite{gk}. 
This is potentially a good thing since the transfer matrix 
derived in \cite{gk} has a number of undesirable features for $m>2$.
We hope to analyze the problem further in the future. 
It is most likely intimately connected with the relation 
between the proper time and the geodesic distance. 
It is clear that the geodesic distance can 
be defined and used as a proper time parameter 
in the context of dynamical triangulations. 
In the case of pure two-dimensional gravity
this choice coincides with the proper time 
defined by the transfer matrix via the fundamental 
deformation $\del \Psi^\dagger (l)$ as discussed, and 
the geometrical reason for this is clear. In other 
cases the relation is less clear, and for instance 
in the Ising model we saw that the basic 
deformation $\del^{\rm new} \Psi^\dagger_0 (l)$ had 
already lost its simple geometric interpretation.
At least the precise relation between the dimension
of proper time and the dimension of geodesic distance 
should be understood. We do not presently 
know the internal Hausdorff dimension of our $(p,q)$ quantum 
universes and consequently not the dimension of geodesic 
distance, but we can relate the dimension of $\cH$ and 
therefore that of the proper time $T$, 
$\dim T = - \dim \cH$
to the choice of conformal background $v_n$. 
The length of the boundary $L$ is connected to the 
variable $\zeta$ via Laplace transformations like \rf{LapPhiPsi},
i.e. 
$\dim \zeta = - \dim L$. 
For a $(p,q)$ model 
the variable $z = \zeta^{1/p}$ and 
$\dim X = \dim \bW^{(p+1)}_{-p} = \dim z^{-p^2}$,
see sec.\ 2.3, sec.\ 5 and appendix 3.
From \rf{pqXn} and \rf{pqvev} we conclude that 
$\dim \sqrt{G} = \dim z^h$, and we finally get 
\beq\label{dimensionT}
\dim T = - \dim \cH = 
-\dim \Bigl(G^{\frac{p-1}{2}} \; X \Bigr) =
\dim L^{\frac{(p-1)(h-p)}{p}-1} \, .
\eeq
For pure gravity $h=5$, see \rf{valu}, 
then eq.\ \rf{dimensionT} indeed reproduces 
$\dim T = \dim L^{1/2}$. 

We have shown by explicit construction 
how the $W^{(2)}$ and $W^{(3)}$ constraints appear 
in the context of the Hamiltonian formalism in pure gravity 
and argued that 
they generalize to any $(p,q)$ model  coupled to 
two-dimensional quantum gravity. 
It is a pleasant surprise that 
the Schwinger-Dyson equations determined the form of $\Hj$ to be 
\beq\label{final1}
\Hj \ = \ -p^{(p-1)}G^{(p-1)/2} X + Y ,~~~~~
X \ = \ \bW^{(p+1)}_{-p} - \dots\dots ,
\eeq
but even more surprising that the vacuum condition 
seems to organize the structure of $X$ as 
\beq\label{final2} 
  X \ = \ - \, \oint \! \frac{d z}{2 \pi i} \oint \! \frac{d s}{2 \pi i} \,
               s^{-p-2} \, \Norder \exp [ - \bW (z,s) ] \Norder \, .
\eeq
Nothing is known about this algebraic structure which seems 
to organize the vacuum of two-dimensional quantum gravity 
coupled to matter. 

We have relied on the Schwinger-Dyson equations as derived 
from the matrix models as a guiding principle for
deriving the form of the Hamiltonian. It is an 
interesting question to which extension it is possible 
to derive the $W$-constraints entirely within a
Hamiltonian context. For instance, is 
it possible to derive $\bW^{(1)}_n Z_f[j]=0$ from $\Hj Z =0$ ?

\vspace{36pt}
\noindent
{\Large \bf Acknowledgment}

We are grateful to R.\ Nakayama for numerous fruitful discussions
and a critical reading of the manuscript.
One of the authors (Y.W.) would in addition like to thank 
M.\ Fukuma and  Y.\ Matsuo  for useful discussions. 
One of the authors (J.A.) acknowledges 
the support of the Professor Visitante Iberdrola Grant
and the hospitality at the University of Barcelona, 
where part of this work was done.

\vspace{36pt}
\noindent
{\Large \bf Appendix 1}

\bea
\delta^{\hbox{\scriptsize new}}\Psi^\dagger_0 (x)
\hspace{-6pt} &=\hspace{-8pt}&
\Biggl[ \, 
x^3 \Bigl( \Psi^\dagger_0 (x) \Bigr)^3 \, 
+ \, ( \frac{\k' x^2}{\k} - 2x + 2\k ) \Bigl( \Psi^\dagger_0 (x) \Bigr)^2 \, 
\nn
&&\phantom{\Biggl[ \, }
+ \, ( \frac{\k^2}{x^3} - \frac{2\k}{x^2} - \frac{\k'+1}{x}
       + \frac{\k'^3}{\k} - \frac{\k'}{\k} - \k + x ) 
     \Psi^\dagger_0 (x) \, 
- \, \k \frac{\prt \Psi^\dagger_0 (0)}{\prt x} x \Psi^\dagger_0 (x) 
\nn
&&\phantom{\Biggl[ \, }
+ \, 2 g \Bigl\{ x 
     \Bigl[ x^2 \Psi^\dagger_0 (x) 
            \bigl( - x \frac{\prt}{\prt x} \Psi_0 (\frac{1}{x}) \bigr)
     \Bigr]^{(+)} \Psi^\dagger_0 (x) \,
\nn
&&\phantom{\Biggl[ \, + \, 2g \Bigl\{ }
     + \, x^3 \Bigl( \Psi^\dagger_0 (x) \Bigr)^2 
            \bigl( - x \frac{\prt}{\prt x} \Psi_0 (\frac{1}{x}) \bigr)
         \Bigr\}
\nn
&&\phantom{\Biggl[ \, }
+ \, 2 g ( \frac{\k' x^2}{\k} - 2x + 2\k ) \Psi^\dagger_0 (x)
     \bigl( - x \frac{\prt}{\prt x} \Psi_0 (\frac{1}{x}) \bigr)
\nn
&&\phantom{\Biggl[ \, }
+ \, 4 g^2 x^3 \Psi^\dagger_0 (x)
     \bigl( - x \frac{\prt}{\prt x} \Psi_0 (\frac{1}{x}) \bigr)^2
\, \Biggr]^{(+)} \,  .
\label{IsingDeform5}
\eea

\vspace{36pt}
\noindent
{\Large \bf Appendix 2}

\vspace{12pt}
\noindent
The explicit form of the Hamiltonian \rf{IsingHgenModeW} is
\bea
\cH \ &=&
  - \, \frac{1}{3} \sum_{a=b+c+d+9}
       \phi^\dagger_d \phi^\dagger_c \phi^\dagger_b a \phi_a \,
  + \, \frac{1}{2} \sum_{n=4}^\infty \sum_{c+d=3n-9 \atop a=b+3n}
       \phi^\dagger_d \phi^\dagger_c \phi^\dagger_b a \phi_a
\nn
&&- \, \frac{G}{2} \sum_{a+b=c+d+9}
       \phi^\dagger_d \phi^\dagger_c b \phi_b a \phi_a \,
  + \, G \sum_{n=2}^\infty \sum_{b+3n-9=d \atop a=c+3n}
       \phi^\dagger_d \phi^\dagger_c b \phi_b a \phi_a \,
\nn
&&   + \, \frac{G}{4} \sum_{n=4}^\infty \sum_{c+d=3n-9 \atop a+b=3n}
       \phi^\dagger_d \phi^\dagger_c b \phi_b a \phi_a
 - \, \frac{G^2}{3} \sum_{a+b+c=d+9}
       \phi^\dagger_d c \phi_c b \phi_b a \phi_a \,
\nn
&&  + \, \frac{G^2}{2} \sum_{n=1}^\infty \sum_{c+3n-9=d \atop a+b=3n}
       \phi^\dagger_d c \phi_c b \phi_b a \phi_a
\nn
&&- \, v_7 \, \Bigl( \,
         \sum_{a=b+c+2}
         \phi^\dagger_c \phi^\dagger_b a \phi_a \,
    - \, \sum_{n=1}^\infty \sum_{a=b+3n}
         \phi^\dagger_{3n-2} \phi^\dagger_b a \phi_a \,
\nn
&&\phantom{- \, v_7 \, \Bigl( \,}
    - \, \frac{1}{2} \sum_{n=1}^\infty \sum_{a+b=3n}
         \phi^\dagger_b \phi^\dagger_a (3n+2) \phi_{3n+2} \, \Bigr)
\nn
&&- \, v_1 \, \Bigl( \,
         \sum_{a=b+c+8}
         \phi^\dagger_c \phi^\dagger_b a \phi_a \,
    - \, \sum_{n=3}^\infty \sum_{a=b+3n}
         \phi^\dagger_{3n-8} \phi^\dagger_b a \phi_a \,
\nn
&&\phantom{- \, v_1 \, \Bigl( \,}
    - \, \frac{1}{2} \sum_{n=1}^\infty \sum_{a+b=3n}
         \phi^\dagger_b \phi^\dagger_a (3n+8) \phi_{3n+8} \, \Bigr)
\nn
&&- \, G v_7 \, \Bigl( \,
         \sum_{a+b=c+2}
         \phi^\dagger_c b \phi_b a \phi_a \,
    - \, \frac{1}{2} \sum_{n=1}^\infty \sum_{a+b=3n}
         \phi^\dagger_{3n-2} b \phi_b a \phi_a \,
\nn
&&\phantom{- \, G v_7 \, \Bigl( \,}
    - \, \sum_{n=0}^\infty \sum_{a+3n=b}
         \phi^\dagger_b a \phi_a (3n+2) \phi_{3n+2} \, \Bigr)
\nn
&&- \, G v_1 \, \Bigl( \,
         \sum_{a+b=c+8}
         \phi^\dagger_c b \phi_b a \phi_a \,
    - \, \frac{1}{2} \sum_{n=3}^\infty \sum_{a+b=3n}
         \phi^\dagger_{3n-8} b \phi_b a \phi_a \,
\nn
&&\phantom{- \, G v_1 \, \Bigl( \,}
    - \, \sum_{n=-2}^\infty \sum_{a+3n=b}
         \phi^\dagger_b a \phi_a (3n+8) \phi_{3n+8} \, \Bigr)
\label{IsingH4}
\\
&&- \, v_7^2 \, \Bigl( \,
         \sum_{a+5=b} \phi^\dagger_b a \phi_a \,
    - \, \sum_{n=1}^\infty \phi^\dagger_{3n+4} (3n-1) \phi_{3n-1}
  \, \Bigr)
\nn
&&- \, 2 v_7 v_1 \, \Bigl( \,
         \sum_{a=b+1} \phi^\dagger_b a \phi_a \,
    - \, \sum_{n=1}^\infty \phi^\dagger_{3n-2} (3n-1) \phi_{3n-1}
  \, \Bigr)
\nn
&&- \, v_1^2 \, \Bigl( \,
         \sum_{a=b+7} \phi^\dagger_b a \phi_a \,
    - \, \sum_{n=1}^\infty \phi^\dagger_{3n-2} (3n+5) \phi_{3n+5}
  \, \Bigr)
\nn
&&- \, G^3 \Bigl\{ \,
            2 ( \phi_1 + \frac{v_1}{G} )^3 \phi_6 \,
      + \, 12 ( \phi_1 + \frac{v_1}{G} )^2 \phi_3 \phi_4 \,
      + \,  9 ( \phi_1 + \frac{v_1}{G} ) \phi_2 \phi_3^2 \,
      + \,  8 \phi_2^3 \phi_3 \, \Bigr\}
\nn
&&- \, \frac{G}{3} \sum_{n=1}^\infty
       \phi^\dagger_{3n} (3n+9) \phi_{3n+9} \,
  - \, 6G^2 \phi_3 \phi_6 \, .
\nonumber
\eea

\vspace{36pt}
\noindent
{\Large \bf Appendix 3}

\vspace{12pt}
\noindent

In this appendix we show that 
the number of fields which describe the Hamiltonian can be reduced by one,  
i.e. the Hamiltonian for ($p$,$q$) model is expressed by $p-1$ fields. 
We have explicitly checked this aspect for $p=2,3,4$ cases. 
We here introduce the following new fields:
\bea
 &&\tilde{\a}^{[i]}(z) \ \define \ \a^{[i]}(z) - \a^{[0]}(z) \, ,
\nn
 &&\a^{[i]}(z) \ = \ \sum_{n \in \dbl{Z}} \a^{[i]}_n z^{-n-1} \, .
\label{tildeJ}
\eea
The current $\a^{[i]}(z)$ is written as 
\beq\label{jstari}
 \a^{[i]\star}(z) \ = \ 
 \frac{1}{\sqrt{G}} \Bigl( v^{[p-i]}(z) + \phi^{[i]\dagger}(z) \Bigr) 
 + \sqrt{G} \frac{\prt}{\prt z} \Bigl( z \phi^{[p-i]}(z) \Bigr) \, , 
\eeq
where 
\bea
 &&\phi^{[i]\dagger}(z) \ = \ \sum_{n=1}^\infty \phi_n^{[i]\dagger} z^{-n-1} , 
 \qquad
   v^{[i]}(z) \ = \ \sum_{n=1}^\infty v^{[i]}_n z^{n-1} , 
\nn
 &&\phi^{[i]}(z) \ = \ \sum_{n=1}^\infty \phi^{[i]}_n z^{n-1} .
\label{phizi}
\eea
Here $\phi_n^{[i]\dagger}$, $\phi_n^{[i]}$ and so on are defined 
as the same as $\a_n^{[i]}$. 
The $X$ operator is expressed by using this $\tilde{\a}^{[i]}$, 
for example, 
\beq\label{Xp2}
 X \ = \ 
 - \, \frac{1}{4} \, \intz 
 : \frac{1}{3z^2} \tilde{\a}^{[1]}{}^3 \, + \, 
   \frac{1}{4z^4} \tilde{\a}^{[1]} :
\eeq
for $p=2$ case,
\beq\label{Xp3}
 X \ = \ 
 - \, \frac{1}{108} \, \intz 
 : \frac{1}{4z^6} \Bigl\{ 
   5 ( \tilde{\a}^{[1]}{}^4 + \tilde{\a}^{[2]}{}^4 )
 - 4 \tilde{\a}^{[1]} \tilde{\a}^{[2]}
   ( \tilde{\a}^{[1]}{}^2 + \tilde{\a}^{[2]}{}^2 ) \Bigr\}
  - \frac{1}{z^8}
   ( \tilde{\a}^{[1]}{}^2 + \tilde{\a}^{[2]}{}^2 ) :
\eeq
for $p=3$ case, and 
\bea
 X \ &=&
  - \, \frac{1}{4^4 \cdot 6} \, \intz 
       \frac{1}{z^{12}} :
    \frac{6}{5} ( \tilde{\a}^{[1]}{}^5 + \tilde{\a}^{[3]}{}^5 - 
                  \tilde{\a}^{[2]}{}^5 )
\nn
&&\phantom{- \, \frac{1}{4^4 \cdot 6} \, \intz \frac{1}{z^{12}} : }
  - 2 ( \tilde{\a}^{[1]}{}^3 \tilde{\a}^{[3]}{}^2 + 
        \tilde{\a}^{[1]}{}^2 \tilde{\a}^{[3]}{}^3 )
  - 5 \tilde{\a}^{[2]}{}^2
      ( \tilde{\a}^{[1]}{}^3 + \tilde{\a}^{[3]}{}^3 )
\nn
&&\phantom{- \, \frac{1}{4^4 \cdot 6} \, \intz \frac{1}{z^{12}} : }
  + 3 \tilde{\a}^{[2]}{}^2
      ( \tilde{\a}^{[1]}{}^2 \tilde{\a}^{[3]} + 
        \tilde{\a}^{[1]} \tilde{\a}^{[3]}{}^2 )
  + 4 \tilde{\a}^{[2]}{}^3 \tilde{\a}^{[1]} \tilde{\a}^{[3]} :
\nn
&&
  - \, \frac{1}{4^4 \cdot 6} \, \intz 
       \frac{1}{z^{14}} :
    2 ( \tilde{\a}^{[1]}{}^3 + \tilde{\a}^{[3]}{}^3 )
  - 3 \tilde{\a}^{[2]}{}^3 
  - 6 \tilde{\a}^{[1]} \tilde{\a}^{[3]} \tilde{\a}^{[2]} :
\nn
&&
  + \, \frac{9}{4^6} \, \intz 
       \frac{1}{z^{16}} \tilde{\a}^{[2]}
\label{Xp4}
\eea
for $p=4$ case.

\end{document}